\documentclass[prb,longbibliography,showpacs,twocolumn,superscriptaddress,amsmath,assume,verbatim]{revtex4-2}
\usepackage{amsmath,amssymb,amsfonts,stmaryrd,wasysym,graphicx, multirow,textcomp, subfigure}
\usepackage{url}
\usepackage[colorlinks=true,citecolor=blue,urlcolor=blue]{hyperref}
\usepackage{romannum}
\usepackage{mathtools}
\usepackage[utf8]{inputenc}
\usepackage{braket}
\usepackage{amsmath}
\usepackage{xcolor}
\usepackage{color,soul} 
\usepackage{bm,xfrac}
\usepackage[normalem]{ulem}

\usepackage{makecell}
\usepackage{tikz}

\usepackage{enumitem}
\usepackage{dcolumn}
\usepackage{bm}
\usepackage[flushleft]{threeparttable}
\usepackage{braket}
\usepackage{array}
\usepackage{booktabs}
\usepackage{float}
\usepackage{setspace}
\hypersetup{colorlinks=true, urlcolor=blue, citecolor=cyan, pdfborder={0 0 0},}

\hypersetup{colorlinks=true, urlcolor=blue, citecolor=cyan, pdfborder={0 0 0}}

\def\be{\begin{equation}}
	\def\ee{\end{equation}}
\def\bea{\begin{eqnarray}}
	\def\eea{\end{eqnarray}}

\begin{document}
\title{Non-Bloch band theory of sub-symmetry-protected topological phases}

	\begin{abstract}
        Bulk-boundary correspondence (BBC) of symmetry-protected topological (SPT) phases relates the non-trivial topological invariant of the bulk to the number of topologically protected boundary states. Recently, a finer classification of SPT phases has been discovered, known as sub-symmetry-protected topological (sub-SPT) phases [\href{https://www.nature.com/articles/s41567-023-02011-9}{Wang \textit{et.al.}, Nature Physics \textbf{19}, 992–998 (2023)}]. In sub-SPT phases, a fraction of the boundary states is protected by the sub-symmetry of the system, even when the full symmetry is broken. While the conventional topological invariant derived from the Bloch band is not applicable to describe the BBC in these systems, we propose to use the non-Bloch topological band theory to describe the BBC of sub-SPT phases. Using the concept of the generalized Brillouin zone (GBZ), where Bloch momenta are generalized to take complex values, we show that the non-Bloch band theory naturally gives rise to a non-Bloch topological invariant, establishing the BBC in both SPT and sub-SPT phases. In a one-dimensional system, we define the winding number, whose physical meaning corresponds to the reflection amplitude in the scattering matrix. Furthermore, the non-Bloch topological invariant characterizes the hidden intrinsic topology of the GBZ under translation symmetry-breaking boundary conditions. The topological phase transitions are characterized by the generalized momenta touching the GBZ, which accompanies the emergence of diabolic or band-touching points. Additionally, we discuss the BBCs in the presence of local or global full-symmetry or sub-symmetry-breaking deformations.
	\end{abstract}

\author{Sonu Verma}
\email{sonu.vermaiitk@gmail.com}
\affiliation{Center for Theoretical Physics of Complex Systems, Institute for Basic Science (IBS) Daejeon 34126, Republic of Korea}
\author{Moon Jip Park}
\email{moonjippark@hanyang.ac.kr}
\affiliation{Department of Physics, Hanyang University, Seoul, 04763, Republic of Korea}
\affiliation{Research Institute for Natural Science, Hanyang University, Seoul, Korea}
\maketitle

\pagenumbering{arabic}

    \section{Introduction}

Topological phases of matter have been understood and classified using the concepts of symmetry and topology across different spatial dimensions. A fundamental characteristic of symmetry-protected topological (SPT) phases is the bulk-boundary correspondence (BBC), which links the concept of bulk topology to the emergence of robust boundary states \cite{Shinsei_Ryu_SPT_phases_2016}. A paradigmatic example of SPT phases in one dimension is the Su-Schrieffer-Heeger (SSH) model \cite{SSH_1979, SSH_1980}. This model consists of two sublattice degrees of freedom with nearest-neighbor hopping terms and respects chiral symmetry. In this case, a $\mathbb{Z}$-valued winding number characterizes the topological boundary modes under open boundary conditions. The BBC in one spatial dimension has been extensively investigated and observed experimentally on various engineered platforms such as optical lattices \cite{Bloch_optical_Zak_exp_2013}, acoustic crystals \cite{Xiao_acoustic_2015}, photonic lattices \cite{Malkova_photonic_2009, Chan_photonic_2014, Gao_photonic_2015}, and topoelectric circuits~\cite{Helbig_topoelectric_2020}.

In the presence of symmetry-breaking perturbations, the topological invariant of the SPT phase is nullified. However, recent studies have identified two types of anomalies: (i) those that destroy the topological invariant while preserving the boundary states \cite{Poli_2017_partial_chiral}, or vice versa, and (ii) those that do not support the existence of edge states while preserving the topological invariant \cite{Longhi_2018_Zak, Longhi_2021_Zak}. The former case has been investigated experimentally, where the topological protection of the edge states has been observed via less stringent sub-symmetry (or partial symmetry) of the system, now known as sub-symmetry-protected topological (sub-SPT) phases \cite{Poli_2017_partial_chiral, Wang2023_subsy, SSR_Schneider2023}. Generally, the conventional topological invariant of the bulk using topological Bloch band theory fails to characterize sub-SPT phases. In case (ii), the bulk-boundary correspondence of topological phases with generalized chiral symmetry (where algebraic deformations break conventional chiral symmetry) has been studied extensively in one \cite{ Ni2019_generalized_chiral_3sub, Gen_Chiral_sym_Hatsugai_2021} and two dimensions \cite{hatsugai_generalized_chiral_2012_1, hatsugai_generalized_chiral_2012_2, hatsugai_generalized_chiral_2015_3, hatsugai_generalized_chiral_2016_4}. Generalized chiral symmetry was initially introduced to characterize tilted Dirac fermions in two dimensions \cite{Goerbig_tilted_Dirac_2008, hatsugai_generalized_chiral_2019_5}. The conventional Bloch topological invariant also can not characterize topological phases with conventional or generalized chiral symmetry in the presence of complete chiral symmetry-breaking perturbations. A systematic understanding of BBC for SPT phases under local or global symmetry-breaking deformations or perturbations is still lacking.

Meanwhile, non-Bloch band theory has successfully demonstrated that non-Hermitian systems exhibit different types of BBC: (i) the complex eigenvalue topology of the bulk leads to the non-Hermitian skin effect, where all bulk states localize at one boundary of the system, and (ii) the wave function topology in the generalized Brillouin zone (GBZ) leads to conventional topological boundary modes \cite{Sato_review_topology_2022, Chen_Fang_review_NHSE_2022, Foa_Torres_NHSE_2018, Wang_non_bloch_TI_2018, Murakami_non_bloch_TI_2019, Murakami_non_bloch_TI_2020}. Recent work~\cite{Verma2024} has shown a different kind of BBC in non-Hermitian systems, originating from the intrinsic topology of the GBZ. A topologically non-trivial GBZ emerges due to general boundary conditions that break the translation symmetry of the system. In this case, the topological phase transition is characterized by the generalized momentum touching of the GBZ, which accompanies the emergence of exceptional points.

In this work, we develop the non-Bloch band theory for the sub-SPT phase. Firstly, we find that the non-Bloch band theory provides a natural non-Bloch topological invariant for the sub-SPT phase, which is related to the reflection amplitude in scattering theory. The topological phase transition of the GBZ is characterized by the generalized momenta touching the GBZ, which accompanies the emergence of diabolic or band-touching points. Secondly, we demonstrate that the intrinsic topology of the GBZ successfully establishes the BBC in systems with conventional or generalized symmetry, even in the presence of global and local complete or partial-symmetry-breaking deformations where conventional Bloch band theory fails to provide insights.

    This paper is organized as follows. In Sec.~\ref{sec:model_ham}, we describe the model Hamiltonian. In Secs.~\ref{sec:Bloch_band},~\ref{sec:Bloch_sub_SPT}, and ~\ref{sec:Bloch_mass}, we summarize the spectral and topological properties of different topological phases in the presence and absence of symmetry-breaking deformations. In Sec.~\ref{sec:non_Bloch}, we discuss the non-Bloch band theory, where Sec.~\ref{sec:gbz} and Sec.~\ref{sec:gbz_top} include the method of finding GBZ and its topology, respectively. In Sec.~\ref{sec:GBBC}, we describe the BBCs for different topological phases in the presence and absence of symmetry-breaking deformations. We summarize our results in Sec.~\ref{sec:summary}.
    \\
    \section{Basic information.}\label{sec:basic} 
    \subsection{Model Hamiltonian}\label{sec:model_ham}
\begin{figure}[htbp!]
		\centering
		\includegraphics[width=1\linewidth]{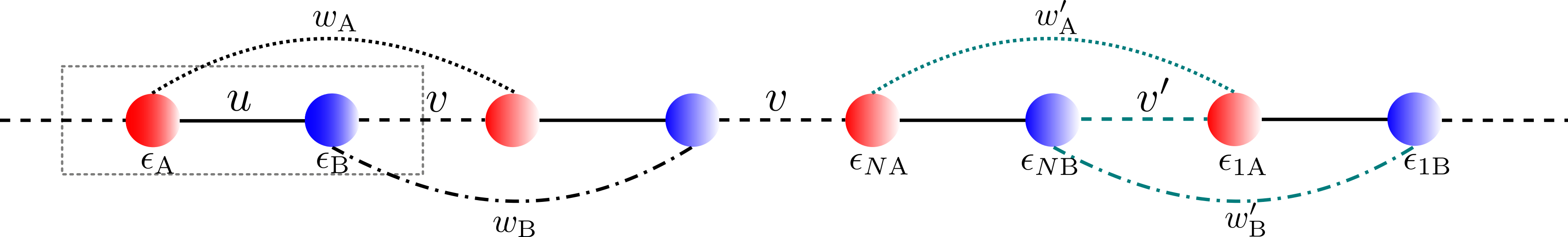}
		\caption{\textbf{Schematics of tight-binding model for one-dimensional lattice with two sublattice degrees of freedom:} Rectangular box shows the unit cell with two sublattice degrees of freedom A, B. $u$ and $v$ denotes the nearest neighbor intra and inter-cell hopping strength. $w_{\rm A}$ ($w_{\rm B}$) denotes the next-nearest neighbor hopping strength for A-A (B-B). $\epsilon_{\rm A}$ ($\epsilon_{\rm B}$) denote the on-site potentials at A (B) sublattice sites in the bulk. $N$ denotes the total number of unit cells. The parameter set $(v', w'_{\alpha},\epsilon_{1\alpha},\epsilon_{N\alpha})$ defines different boundary conditions.}
	\label{fig:tight_binding_GBC}
    \end{figure}

    We consider a generic tight-binding model on a one-dimensional (1D) lattice with two sublattice degrees of freedom [Fig.~\ref{fig:tight_binding_GBC}]. The corresponding model Hamiltonian $\hat{H}= \hat{H}_{\text{SSH}}+\hat{H}_{\text{AA}}+\hat{H}_{\text{BB}}+\hat{H}_{\rm d}$ is described as follows
    \bea\label{eq:main_Ham}
    \hat{H}_{\rm SSH}&=&\sum_{j=1}^{N}
    \big[ u \hat{c}^\dagger_{j \rm B}\hat{c}_{j \rm A}
    +v(1-\delta_{j N}) \hat{c}^\dagger_{(j+1) \rm A}\hat{c}_{j \rm B} + {\rm h.c.}\big],
    \nonumber\\
    \hat{H}_{\alpha\alpha}&=&\sum_{j=1}^{N}\big[
     \epsilon_{\alpha} \hat{c}^\dagger_{j\alpha}\hat{c}_{j\alpha}
    + w_{\alpha} (1-\delta_{j N}) \hat{c}^\dagger_{(j+1)\alpha}\hat{c}_{j\alpha}
    +{\rm h.c.}\big],
    \nonumber\\
     \hat{H}_{\rm d}&=&v'\hat{c}^\dagger_{1\rm A}\hat{c}_{N \rm B}
     +\sum_{\alpha}w'_{ \alpha}\hat{c}^\dagger_{1\alpha}\hat{c}_{N\alpha}+{\rm h.c.}\nonumber\\
     &+&\sum_{\alpha}[\epsilon_{1\alpha} \hat{c}^{\dagger}_{1\alpha}\hat{c}_{1 \alpha}+\epsilon_{N\alpha} \hat{c}^{\dagger}_{N\alpha}\hat{c}_{N \alpha}],
    \eea
    here $\hat{c}^{\dagger}_{j\alpha} (\hat{c}_{j\alpha})$ denote the creation (annihilation) operator at $j$-th unit cell with sublattice degrees of freedom $\alpha\in\lbrace {\rm A},~{\rm B}\rbrace$. $u$, $v$ denote the nearest neighbor intracell and intercell hopping. $\epsilon_{\rm A (B)}$ denotes the on-site potential at the sublattice site A (B). $N$ is the total number of unit cells. $w_{\rm A (B)}$ denotes the next-nearest neighbor intercell hoppings between A-A (B-B) sites. $\hat{H}_{\rm SSH}$ represents the model Hamiltonian for the conventional 1D Su-Schriefer-Hegger (SSH) model. $\hat{H}_{\rm \alpha\alpha}$ and $\hat{H}_{\rm d}$ can be responsible for the breaking of certain symmetries of $\hat{H}_{\rm SSH}$ and therefore represent the bulk (global) and the boundary (local) deformation term, respectively. 
    $\hat{H}_{\rm d}$ characterizes different generalized boundary conditions (see Appendix \ref{supp:sec:def_GBC} for more details). For example, the parameter sets $(v',w'_{\alpha},\epsilon_{1\alpha},\epsilon_{N\alpha}) =(v,w_{\alpha},0,0)$ and $(v',w'_{\alpha},\epsilon_{1\alpha},\epsilon_{N\alpha}) =(0,0,0,0)$ define the periodic boundary condition (PBC) and the open boundary condition (OBC), respectively. 
    \\
    \subsection{Topological Bloch band theory}\label{sec:Bloch_band}
    We consider the periodic boundary condition (PBC). In this case, the model Hamiltonian can be diagonalized concerning the associated conserved Bloch momenta $k$, which are confined within the first Brillouin zone (BZ): $\lbrace k,~k\in(-\pi,\pi]\rbrace.$ Accordingly, with the help of the real space field operators $\hat{c}_{j\alpha}=\frac{1}{\sqrt{N}}\sum_{k}e^{i k j}\hat{c}_{k\alpha},~\alpha\in\lbrace\rm A,~B\rbrace$, the Hamiltonian takes the following form
    $\hat{H}=\sum_{k\in \rm BZ}\sum_{ \alpha,\beta\in\lbrace \rm A,B\rbrace}\hat{c}_{k\alpha}^{\dagger}(H(k))_{\alpha\beta}\hat{c}_{k\beta}.$
    The periodic Bloch Hamiltonian $H(k)$ is expressed as follows,
    \begin{align}\label{eqn:ham_Bloch2}
        H(k)&=d_0(k)\sigma_0+\textbf{d}(k)\cdot \bm{\sigma},~d_0(k)=\epsilon^{+}+w^{+}\cos(k),\nonumber\\
        \textbf{d}(k)&=(u+v\cos(k),~v\sin(k),~\epsilon^{-}+w^{-}\cos(k)),
    \end{align}
     here $\textbf{d}(k)=(d_x(k),d_y(k),d_z(k))$ is a cartesian real vector and $\bm{\sigma}=(\sigma_x,\sigma_y,\sigma_z)$ is matrix valued vector, $\epsilon^{\pm}=(\epsilon_{\rm A}\pm \epsilon_{\rm B})/2$, $w^{\pm}=w_{\rm A}\pm w_{\rm B}$. $\sigma_{0}$ and $\sigma_{j=x,y,z}$ denote the $2\times 2$ identity and Pauli matrices, respectively, acting on two sublattice degrees of freedom $\lbrace{\rm A, B}\rbrace$. The energy eigenvalues and corresponding eigenstates are expressed as follows
     \begin{align}
         E_{\pm}(k)&=d_0(k)\pm \sqrt{|\textbf{d}(k)|},\nonumber\\
         |\Psi_{\pm}(k)\rangle&=\frac{1}{\sqrt{2|\textbf{d}(k)|(|\textbf{d}(k)|\pm d_z(k))}}\begin{pmatrix}
             d_z(k)\pm |\textbf{d}(k)|\\
             d_x(k)+i d_y(k)
         \end{pmatrix}.
     \end{align}
    In the absence of sublattice on-site potentials and next-nearest neighbor hopping, the model reduces to the conventional SSH model with the Bloch Hamiltonian $H_{\rm SSH}(k) = d_x(k)\sigma_x + d_y(k)\sigma_y$.
    This model belongs to the BDI class of 10-fold classification of symmetry-protected topological phases as it respects the time-reversal ($\mathcal{T}=\mathcal{K}$) [$H_{\rm SSH}^*(k)=H_{\rm SSH}(-k)$], particle-hole ($\mathcal{P}=\sigma_z \mathcal{K}$) [$\sigma_z H_{\rm SSH}^*(k)\sigma_z^{-1}=-H_{\rm SSH}(-k)$], and chiral symmetry [$\sigma_z H_{\rm SSH}(k)\sigma_z^{-1}=-H_{\rm SSH}(k)$] \cite{Hughes_2021}. The chiral symmetry ensures that a state $|\psi_{+}(k)\rangle$ with energy $E_{+}(k)$ guarantees a chiral partner states $|\psi_{-}(k)\rangle=\sigma_z|\psi_{+}(k)\rangle$ with energy $E_{-}(k)=-E_{+}(k)$. Consequently, the topological phases are characterized by the strong topological invariant, specifically known as the Bloch winding number $\mathcal{W}_{\rm Bloch}$, which is defined as winding of the off-diagonal element of Bloch Hamiltonian \cite{Shinsei_Ryu_SPT_phases_2016}, $\mathcal{W}_{\rm Bloch}=\oint_{\rm BZ} \partial_{k}\log[d_x(k)-id_y(k)]~d k/(2\pi i)$.
    The system has a trivial phase ($\mathcal{W}_{\rm Bloch}=0$) for $|u|>v$ and a topologically non-trivial phase ($\mathcal{W}_{\rm Bloch}=1$) for $|u|<v$. Accordingly, consistent with the bulk-boundary correspondence, in the topologically non-trivial phase, one observes topologically protected zero-energy boundary modes in the semi-infinite SSH chain. In this work, we show that the generalized Brillouin zone topology also establishes the BBC for SPT phases [see Secs.~\ref{sec:gbz_top},~\ref{sec:BBC_SPT}].

    Now, we consider the case of finite but equal on-site potentials on both sublattice sites ($\epsilon_{\rm A}=\epsilon_{\rm B}$) and next-nearest neighbor hoppings ($w_{\rm A}=w_{\rm B}$). In this case, the Bloch Hamiltonian reduces to $H(k) = d_0\sigma_0+d_x\sigma_x + d_y\sigma_y$
    which naturally breaks the conventional chiral symmetry [$\sigma_z H(k)\sigma_z^{-1}\neq -H(k)$]. However, the system still respects the inversion symmetry: $\sigma_x H(k)\sigma_x^{-1}=H(-k)$ and the generalized chiral symmetry: $\gamma(k)^{\dagger} H(k)\gamma(k)=-H(k)$ with $\gamma(k)=e^{\theta(\sigma_x+\sigma_y)/2}\sigma_z e^{-\theta(\sigma_x+\sigma_y)/2}, ~\tanh{[\theta\sqrt{2}]}=d_0(k)\sqrt{2}/[d_x(k)+d_y(k)]$ \cite{ Ni2019_generalized_chiral_3sub, Gen_Chiral_sym_Hatsugai_2021}. Adding a constant term to the Hamiltonian does not change the Bloch eigenstates. Therefore, the conventional Bloch winding number remains unchanged to this deformation and related to the Zak phases as \cite{Zak_phase_1989} $\phi_{\pm}=i\oint_{\rm BZ} \langle\Psi_{\pm}(k)|\partial_k\Psi_{\pm}(k)\rangle d k =\pi \mathcal{W}_{\rm Bloch}$. The topologically non-trivial phases are characterized by quantized $\pi$ Zak phase (mod $2\pi$) and correspond to the boundary modes in the system. However, varying the next nearest neighbor hopping $w_{\rm A}=w_{\rm B}$, the band gap closes at $w_{\rm A}=v/2$, and the boundary modes disappear, whereas the Zak phase and the winding number are still quantized \cite{Longhi_2018_Zak, Longhi_2021_Zak}. Therefore, in this case, the conventional Bloch topological invariants fail to establish the BBC. This work shows that the generalized Brillouin zone topology successfully establishes the BBC for generalized chiral symmetric phases [see Sec. \ref{sec:bbc_gen_chiral}].
    
    \begin{figure*}[t!]
		\centering
		\includegraphics[width=1\linewidth]{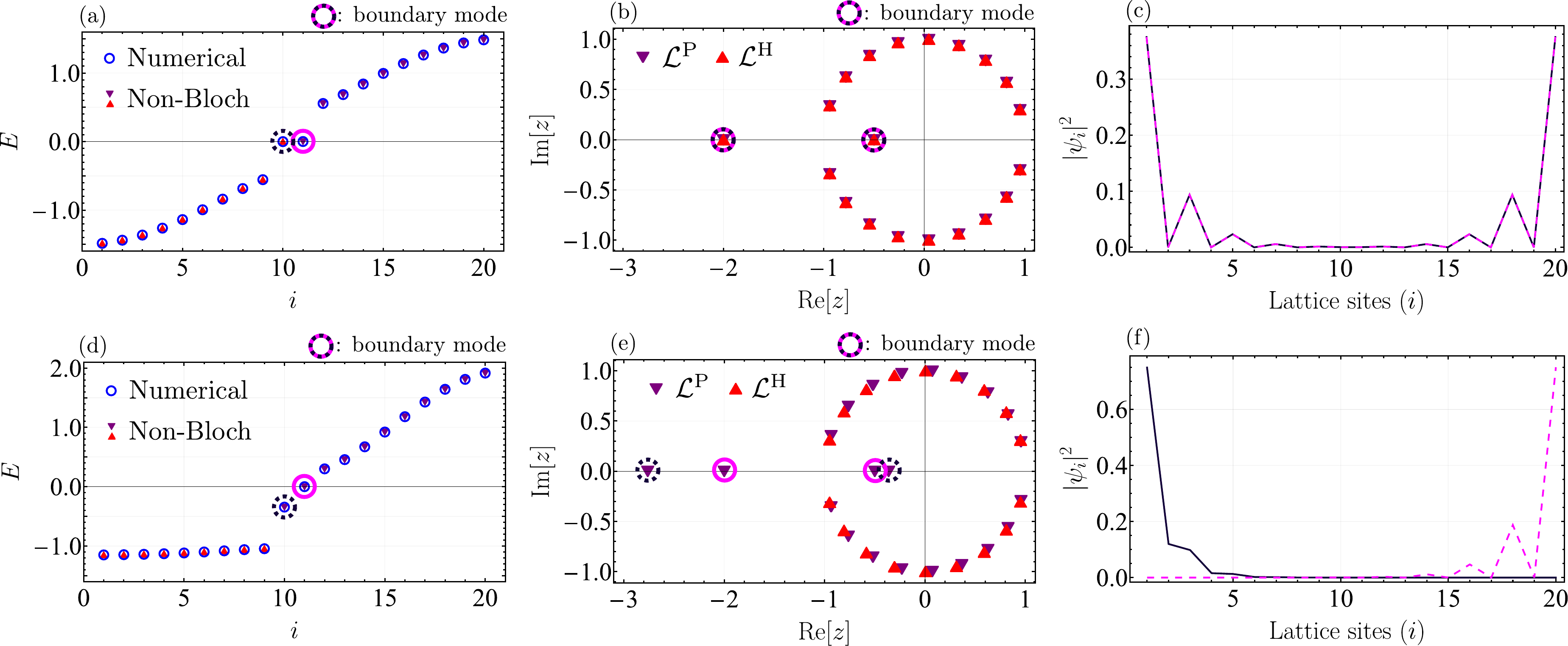}
		\caption{\textbf{Spectral properties in the open boundary condition:} [(a) and (d)], [(b) and (e)], and [(c) and (f)] show the eigenvalue spectra, generalized momenta, and wave function amplitudes for boundary modes, respectively, for the [symmetry-protected topological phases and sub-symmetry-protected topological phases]. Circles mark the boundary modes, and $\mathcal{L}^{\rm P}$ ($\mathcal{L}^{\rm H}$) denote the generalized Brillouin zones for particle and hole sectors. The eigenvalue spectra obtained via exact numerical diagonalization and non-Bloch band theory coincide. Parameters: (a), (b), and (c): $u=0.5,~w_{\rm A}=0.0$. (d), (e), and (f): $u=0.5,~w_{\rm A}=0.4$. All panels have $w_{\rm B}=\epsilon_{\alpha}=\epsilon_{1\alpha}=\epsilon_{N\alpha}=v'=w'_{\alpha}=0,\alpha\in\lbrace \rm A, B\rbrace,~N=10.$}
	\label{fig:eig_gbz_amp}
    \end{figure*}
    
    \subsection{Sub-symmetry-protected topological phases.}\label{sec:Bloch_sub_SPT} 
    The sub-symmetry considers perturbations that break the original symmetry but preserve a less strict sub-symmetry \cite{Wang2023_subsy}. For example, we consider the chiral symmetry, defined as $\lbrace S, H\rbrace =0$. Here $S=P_{\rm A}-P_{\rm B}$, with $P_{\rm A}$ ($P_{\rm B}$) being the projection operators on A (B) sublattice in the real or reciprocal space and $H$ denotes the matrix representation of the model Hamiltonian. The chiral symmetry is equivalent to two possible sub-symmetries of the system. Namely, the A sub-symmetry and the B sub-symmetry are defined as follows \cite{Wang2023_subsy}
    \bea
     \Sigma_z H \Sigma_z^{-1} P_{\alpha} =-H P_{\alpha},~\alpha=\lbrace {\rm A},~{\rm B}\rbrace.
    \eea
    The conventional SSH model has both sub-symmetries, guaranteeing a pair of zero-energy topological boundary modes for each sublattice edge. The sub-SPT phases possess global or local deformations that do not break all sub-symmetries of the system. For example, $H = H_{\rm SSH} + H_{\rm AA}$ and $H = H_{\rm SSH} + H_{\rm BB}$ respects the B sub-symmetry and A sub-symmetry, respectively. In the presence of global deformation term $w_{\rm A}\neq 0,~w_{\rm B}=0$, the Bloch Hamiltonian for B sub-symmetry protected phase is expressed as follows
    \begin{align}
        H(k)=\begin{pmatrix}
            \epsilon_{\rm A}+2 w_{\rm A}\cos(k) & u + v e^{-i k}\\
            v+v e^{i k} & 0
        \end{pmatrix}.
    \end{align}
    In this case, a zero energy mode still exists, which is protected by B sub-symmetry \cite{Poli_2017_partial_chiral, Wang2023_subsy}. The conventional Brillouin zone topology or Bloch band theory fails to characterize this zero energy boundary mode and hence can not establish the correct BBC. Establishing the BBC in these systems is the main goal of this work. In Secs.~\ref{sec:gbz_top},~\ref{sec:BBC_sub_SPT1}, and  \ref{sec:BBC_sub_SPT2}, we show that the generalized Brillouin zone topology successfully characterizes sub-SPT phases.
    
   \subsection{Conventional/generalized chiral symmetry breaking by the mass term.}\label{sec:Bloch_mass}
   When the on-site sublattice potential difference characterized by constant mass term ($m=\epsilon^{-}$) and variable mass term ($m=w^{-}\cos(k)$) or next-nearest neighbor hopping exist, Eq.~\ref{eqn:ham_Bloch2} describes the model Hamiltonian with broken chiral symmetry. For $w^{\pm}=0$, the constant mass term only displaces the energy of the boundary modes by $\pm m$. However, for $w^{-}=0$ but finite $w^{+}$, the system respects the generalized chiral symmetry. In this case, in the presence of a mass term, both the energies and corresponding eigenstates of boundary modes become dependent on $m$. Therefore, the boundary modes disappear at some critical value of $m$ where the localization length of the boundary modes diverges \cite{Gen_Chiral_sym_Hatsugai_2021}.
   
   Mass term breaks the chiral and the inversion symmetries of the system, and hence, the quantization of the winding number and the Zak phase is lost. Nevertheless, we show in Sec.~\ref{sec:BBC_mass} and Sec.~\ref{sec:bbc_gen_chiral} that the intrinsic topology of GBZ can still characterize the presence of boundary modes in the system.
    
    \section{Non-Bloch band theory}\label{sec:non_Bloch} 
    \subsection{Energy spectra and generalized Brillouin zones}\label{sec:gbz}
    In this section, we use the non-Bloch band theory to derive the spectral properties of the sub-SPT phase. We consider B-SubSy protected phase ($\hat{H} = \hat{H}_{\rm SSH} + \hat{H}_{\rm AA}$) which can straightforwardly be reduced to the symmetry-protected phase ($\hat{H}_{\rm SSH}$) by substituting $w_{\rm A}=0$ (see Appendices \ref{supp:sec:non_Bloch_theory_g} and \ref{supp:sec:non_Bloch_theory_sub_SPT} for the derivation of non-Bloch band theory in more generic cases).
    We follow the methodology of the non-Bloch band theory developed in Ref.~\cite{chenGBC2021} to solve the eigenvalue problem $\hat{H}|\Psi\rangle=E|\Psi\rangle$ in the open boundary condition (OBC). 
    Since the translation symmetry of the bulk is intact, we can still represent the eigenstates as $|\Psi\rangle =1/\sqrt{\mathcal{N}} \sum_{n=1}^{N}(\psi_{n \rm A}\hat{c}^{\dagger}_{n \rm A}+\psi_{n \rm B}\hat{c}^{\dagger}_{n \rm B})|0\rangle $, where $\psi_{n\alpha}\equiv\psi_{n\alpha}(z_1,z_2)=c_1  \phi_{\alpha}^{(1)}z_1^n +c_2  \phi_{\alpha}^{(2)}z_2^n$ with $\alpha\in\lbrace {\rm A},~{\rm B}\rbrace $ is the wave-function ansatz and $z_\alpha \in \mathbb{C}$ is the generalized complex momenta. Here $\mathcal{N}$ and $N$ denote the normalization constant and total number of unit cells, respectively. The eigenstates are required to be self-consistent with both bulk and boundary equations. Explicitly, the bulk equations are given as,
    \bea 
        \label{eq:bulk_eqn}
            \frac{\phi_{\rm A}^{(j)}}{\phi_{\rm B}^{(j)}}=-\frac{v/z_j+u}{w_{\rm A}/z_j -E + w_{\rm A}z_j}=\frac{E}{v z_j+u},~j=1,~2.
    \eea
    The bulk equation can be rewritten as  a second-order polynomial equation in $z_j$ and admits the complex momenta solutions in pairs, $(z_1,~z_2)$ satisfying  $z_1 z_2=1$, which is given as
    \bea 
        \label{eq:bulk_eq_poly}
	&E^2-w_{\rm A}E\Big(\frac{1}{z_j}+z_j\Big)- \Big(\frac{v}{z_j}+u\Big)(v z_j+u)= 0.
    \eea 
    Therefore, the paired momenta $(z_1,~z_2)$ determines the eigenvalues and the corresponding eigenstates. In addition, the boundary equations are given as,
    \begin{align}
	\label{Eq_bmat}
	M_{\rm b}(c_1,~c_2)^\textrm{T}=0,~ {\rm where}~M_{\rm b}=
	\begin{pmatrix}
	P(z_1) & P(z_2) \\
	Q(z_1) & Q(z_2)
	\end{pmatrix}.
    \end{align}
    with, $P(z_j) = -v\phi_{\rm B}^{(j)}-w_{\rm A} \phi_{\rm A}^{(j)},~Q(z_j) =-v z_j^{N+1}\phi_{\rm A}^{(j)}$ where $\phi_{\rm A}^{(j)}$ and $\phi_{\rm B}^{(j)}$ are related via Eq.~\eqref{eq:bulk_eqn}.  The non-trivial solutions of the boundary equations [Eq.~\eqref{Eq_bmat}] can be obtained by solving the determinant of the boundary matrix, represented as $\det[M_{\rm b}]=0$. We denote these paired momenta solutions as $(z_{1n},z_{2n})=(e^{i k_n},e^{-i k_n})$, where $k_n\equiv k\in\mathbb{C},$ and $n\in\lbrace 1,..., N\rbrace$ satisfying the constraint $z_{1n}z_{2n}=1$ as dictated by the bulk equations [Eq.~\eqref{eq:bulk_eqn}]. Consequently, the condition $\det[M_{\rm b}]=0$ can be expressed as follows,
    \begin{align}\label{eq:sol_phi_restore}
        &u v\sin[(N+1) k]+v^2\sin[N k]\nonumber\\
        &=-E_{\pm}(k) w_{\rm A}\sin[(N+1)k],
    \end{align}
    where the energy eigenvalues $E_{\pm}(k)$ are given by [Eq.~\ref{eq:bulk_eq_poly}]
    \begin{align}
    \label{eq:bulk_eigenvalue}
        E_{\pm}(k)&=w_{\rm A}\cos[k]\nonumber\\
        &\pm\sqrt{(w_{\rm A}\cos[k])^2 + u^2+v^2+2 u v \cos[k]}.
    \end{align}
    The contours formed by the paired (generalized) momenta describe the GBZs $\mathcal{L}_1\times \mathcal{L}_2$, characterizing the bulk states. Generally, Eq.~\ref{eq:sol_phi_restore} yields two distinct sets of non-zero solutions, corresponding to the particle $E_{+}$ and hole $E_{-}$ sectors, respectively. Consequently, we designate the GBZs within the particle (hole) sectors as $\mathcal{L}_{1}^{\rm P}\times\mathcal{L}_{2}^{\rm P}$ ($\mathcal{L}_{1}^{\rm H}\times\mathcal{L}_{2}^{\rm H}$), respectively. Furthermore, the disjoint generalized momenta $z_{2}^{(\rm b)}$ ($z_{1}^{(\rm b)}$) outside (inside) GBZs (the unit circle ($|z_{1}|=|z_{2}|=1$) for Hermitian systems) describes the evanescent wave localized on the left (right) boundary. The localization lengths of these boundary states are determined by $\xi=-1/\log|z_{1}^{({\rm b})}|$. Touchings of the disjoint generalized momenta of the two GBZs correspond to the diabolic points or the band touching points where two eigenvalues coincide, and localization lengths of the boundary modes diverge. 

    \begin{figure*}[htbp!]
		\centering
		\includegraphics[width=1\linewidth]{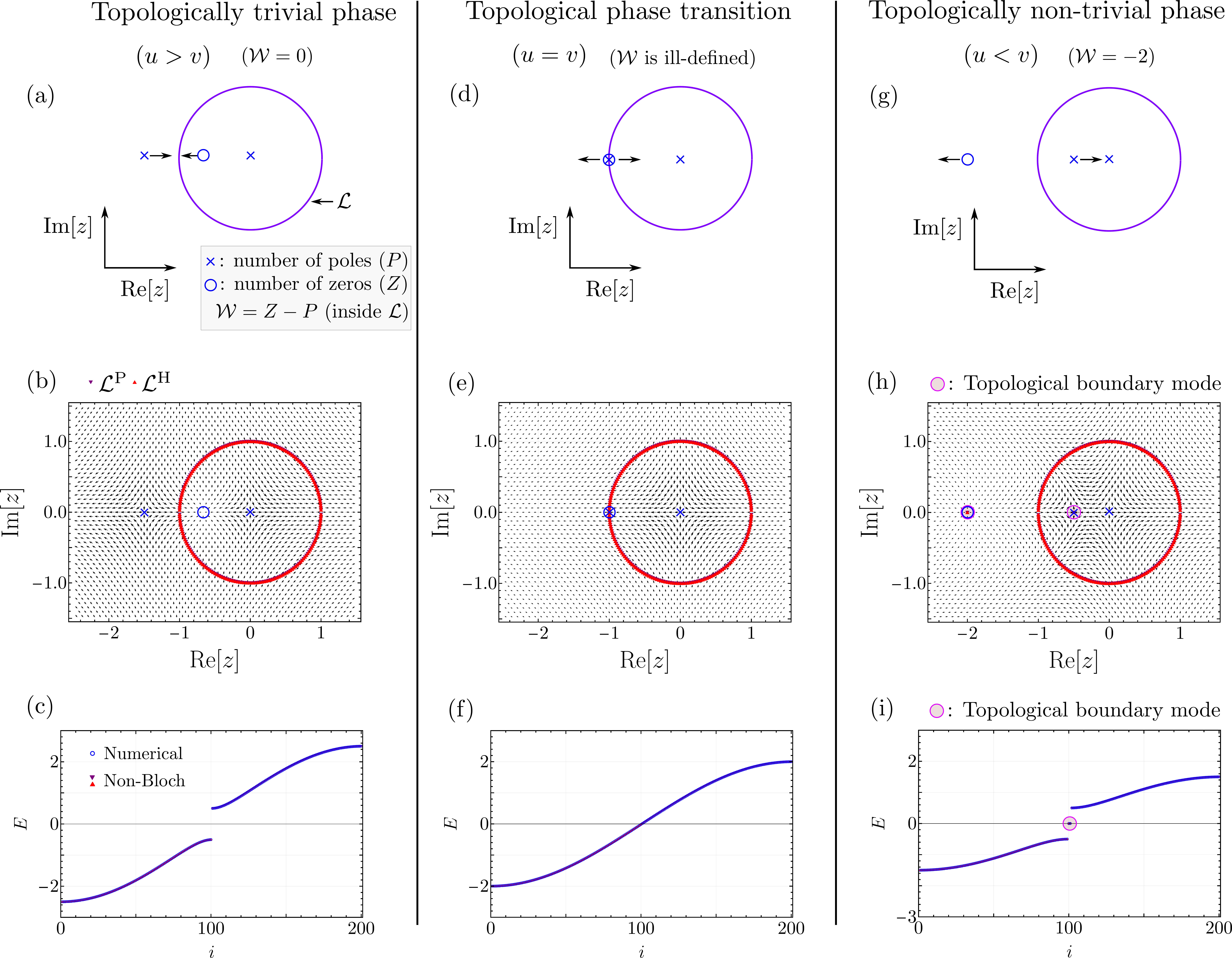}
		\caption{\textbf{Topological phase transition of generalized Brillouin zones for the symmetry-protected topological phase in the open boundary condition.} In this case, the topological phase transitions of the generalized Brillouin zone (GBZ: $\mathcal{L}^{\rm P}=\mathcal{L}^{\rm H}\equiv\mathcal{L}$) are described by the total change in the number of zeros and poles of the meromorphic function $M_{\rm b}^{+}(z)$ inside $\mathcal{L}$ which determine the non-Bloch topological invariant $\mathcal{W}$. [(a), (d), and (g)], [(b), (e), and (h)], and [(c), (f), and (i)] describe the contours of GBZs together with zeros and poles of the meromorphic function $M_{\rm b}^{+}(z)$, finite size sub-symmetry protected GBZs ($\mathcal{L}^{\rm P}, \mathcal{L}^{\rm H}$) together with complex vector plots of the meromorphic function $M_{\rm b}^{+}(z)$, and eigenvalue spectra, respectively, for the [topologically trivial phase ($u>v$), topological phase transition ($u=v$), and topologically non-trivial phase (($u<v$)), respectively. Parameters: [(a), (b), and (c)]: $u=1.5$. [(d), (e), and (f)]: $u=1.0$. [(g), (h), and (I)]: $u=0.5$. All panels have $~v=1.0,~~N=100,~w_{\rm \alpha}=\epsilon_{\alpha}=\epsilon_{1\alpha}=\epsilon_{N\alpha}=v'=w'_{\alpha}=0,~\alpha\in\lbrace \rm A,B\rbrace.$}
	\label{fig:OBC_TPT_SPT}
    \end{figure*}
    
    Fig.~\ref{fig:eig_gbz_amp} (a)-(b) and (d)-(e) show the energy spectra-GBZs of the SPT and sub-SPT phases, respectively. In the SPT phase ($w_{\rm A}=0$), GBZs for particle and hole sectors coincide [See Fig.~\ref{fig:eig_gbz_amp} (b)]. When $u<v$, the bulk energy spectra exhibit particle-hole symmetry, accompanied by two additional zero-energy boundary modes. The topological boundary modes are described by the two degenerate disjoint generalized momenta $(z_1^{\rm (b)},z_2^{\rm (b)})=(u/v,v/u)$ in the complex momenta plane [See Fig.~\ref{fig:eig_gbz_amp} (a)-(b)]. The boundary modes are protected by both A and B sub-symmetries with zero wave function amplitudes at the A and B sublattice sites, respectively [See Fig.~\ref{fig:eig_gbz_amp} (c)]. 
    
    In contrast, due to the absence of the chiral and the particle-hole symmetries, the sub-SPT phase ($w_{\rm A}\neq 0$) exhibits the two separated GBZs in the particle and hole sectors [Fig.~\ref{fig:eig_gbz_amp} (e)]. While the degeneracy of disjoint generalized moments is broken, we still observe a zero energy boundary mode with disjoint generalized momenta $(z_1^{\rm (b)},z_2^{\rm (b)})=(u/v,v/u)$ that is protected by B sub-symmetry [Fig.~\ref{fig:eig_gbz_amp} (d) and (e)]. Consequently, the B sub-symmetry-protected boundary mode exhibits zero wave function amplitude at B sublattice sites, mirroring the behavior observed in the corresponding SPT phase. Conversely, the boundary mode endowed with finite energy manifests finite wave function amplitude at A sublattice sites [see Fig.~\ref{fig:eig_gbz_amp} (e), (f)].

     \begin{figure*}[htbp!]
		\centering
		\includegraphics[width=1.0\linewidth]{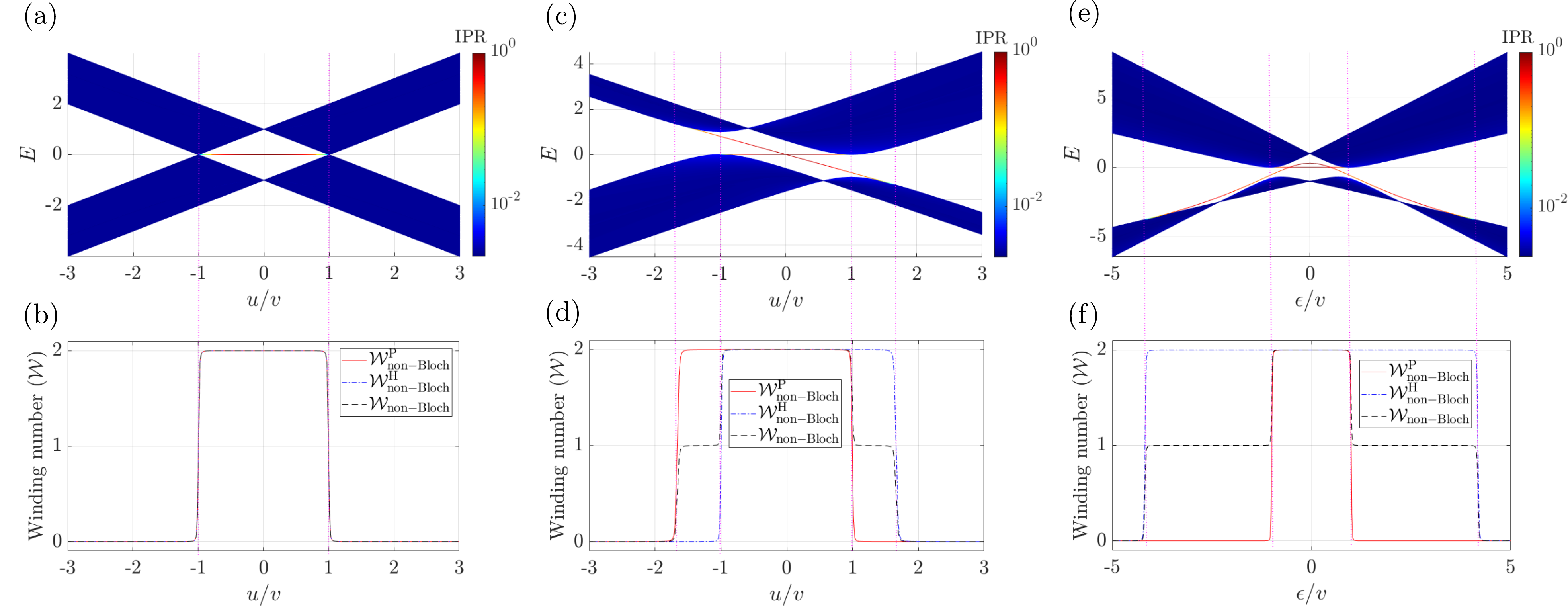}
		\caption{\textbf{Generalized bulk-boundary correspondence in the open boundary condition.} [(a) and (b)], [(c) and (d)], and [(e) and (f)] show [the variation of energy spectra ($E$) and non-Bloch topological invariants ($\mathcal{W}$) with system parameters] in the symmetry-protected topological (SPT) phase, B sub-SPT phase, B sub-SPT phase in the presence of local deformation, respectively. The non-trivial values of $\mathcal{W}_{\rm non-Bloch}$ characterize the existence of the boundary states. IPR denotes the inverse participation ratio, which distinguishes the bulk states (IPR $\sim O(1)$) and boundary states (IPR$\sim O(1/N)$). Parameters: (a), (b): $w_{\rm \alpha}=\epsilon_{\alpha}=\epsilon_{1\alpha}=\epsilon_{N\alpha}=v'=w'_{\alpha}=0$. (c), (d): $w_{\rm A}=0.5,~w_{\rm B}=0,~\epsilon_{\alpha}=\epsilon_{1\alpha}=\epsilon_{N\alpha}=v'=w'_{\alpha}=0$. (e), (f): $u=\epsilon,~w_{\rm A}=0.4\epsilon,~=\epsilon_{1\rm A}=0.3,~\epsilon_{1\rm B}=\epsilon_{N\alpha}=w_{\rm B}=\epsilon_{\alpha}=v'=w'_{\alpha}=0$. Here $\alpha\in\lbrace \rm A,B\rbrace$ and all panels have $v=1,~N=300.$}
	\label{fig:E_W_SPT_sub_SPT}
    \end{figure*}
    
    \subsection{Topology of generalized Brillouin zone}\label{sec:gbz_top}
     In the generalized boundary conditions, the wave functions $\psi_{n\alpha}=c_1  \phi_{\alpha}^{(1)}z_1^n +c_2  \phi_{\alpha}^{(2)}z_2^n,~\alpha=\lbrace\rm A, B\rbrace$ are characterized by multiple GBZs, promoting the expanded spinor $(c_1\phi_{\alpha}^{(1)},c_2\phi_{\alpha}^{(2)})^{\textrm{T}}\equiv(c'_1,c'_2)^\textrm{T}$ representations of the eigenstates. One can think of $c'_1,~c'_2$ as the coefficients of outgoing ($z_1\equiv e^{-i k}$) and incoming waves ($z_2\equiv e^{i k}$) in the full wave function~\cite{Verma2024}. Accordingly, the meromorphic function $M_{\rm b\beta}^{+}(z_1,z_2)\equiv-\frac{c'_1}{c'_2}=\frac{P(z_2)\phi_{\alpha}^{(1)}}{P(z_1)\phi_{\alpha}^{(2)}}$ corresponds to the reflection amplitude for the scattering problem. The existence of the topological boundary modes is detected by the non-trivial winding number associated with the meromorphic function $M_{\rm b\beta}^{+}(z_1,z_2)$. The change of the winding number signifies the topological phase transition. Using Cauchy's argument principle, we can define the winding number $\mathcal{W}_{\rm non-Bloch}^{\beta}$ as follows \begin{align}\label{eq:winding_arg}
        \mathcal{W}_{\pm}^{\beta}&=\frac{1}{2\pi i}\oint_{\mathcal{L}_1^{\beta}\times\mathcal{L}_2^{\beta}}\frac{1}{M_{\rm b\beta}^{\pm}(z_1,z_2)}\frac{d M_{\rm b\beta}^{\pm}(z_1,z_2)}{dz_1} dz_1\nonumber\\
    &= \frac{1}{2\pi}[\textrm{arg}~M_{\rm b\beta}^{\pm}]_{\mathcal{L}_1^{\beta}\times \mathcal{L}_2^{\beta}}=Z-P,
    \end{align}
     where $[\textrm{arg}~M_{\rm b \beta}^{\pm}]_{\mathcal{L}_1^{\rm \beta}\times \mathcal{L}_2^{\rm \beta}}$ is the change of phase of $M_{\rm b \beta}^{\pm}$ as $(z_1,z_2)$ goes along GBZs of sector $\beta={\rm P, H}$. $Z~(P)$ denotes the total number of zeros (poles) inside the GBZs. For the SPT phase in the open boundary condition, we show exemplifies the GBZs with zeros and poles of $M_{\rm b \beta}^{+}$, generalized momenta together with complex vector plot of $M_{\rm b\beta}^{+}$, and energy spectra in [Fig.~\ref{fig:OBC_TPT_SPT} (a),(d), and (g)], [Fig.~\ref{fig:OBC_TPT_SPT} (b), (e), and (f)], and [Fig.~\ref{fig:OBC_TPT_SPT} (g), (h), and (i)], respectively, for the [trivial ($u>v$), critical ($u=v$), and non-trivial ($u<v$) topological phases, respectively]. For $u>v$, GBZs are topologically trivial [$\mathcal{W}=0$, Fig.~\ref{fig:OBC_TPT_SPT} (a)] as it encloses an equal number of poles and zeros of $M_{\rm b \beta}^{+}$. At $u=v$, the non-zero zeros and poles of $M_{\rm b \beta}^{+}$ touch the GBZs and the system undergoes the topological phase transition [Fig.~\ref{fig:OBC_TPT_SPT} (d)]. Finally, for $u<v$, the topologically non-trivial GBZ encircles a pair of poles, which results in the non-trivial value of the winding number [Fig.~\ref{fig:OBC_TPT_SPT} (g)]. In the topologically non-trivial phase [Fig.~\ref{fig:OBC_TPT_SPT} (e) and (f)], the disjoint generalized momenta (marked with circles) characterize the topological boundary modes. In this case, the topological phase transition is characterized by the touching of disjoint generalized momenta to the GBZs, which corresponds to the band touching point [Fig.~\ref{fig:OBC_TPT_SPT} (h) and (i)].
     
     The quantization of the $\mathcal{W}_{\rm non-Bloch}^{\beta}$ is possible due to the presence of the symmetry $\{ \tilde{M}_{\rm b\beta}, \sigma_z \}=0$ of the transformed boundary matrix $\tilde{M}_{\rm b\beta}$ [Eq.~\eqref{Eq_bmat}], which is defined by the eigenvalue equation, $\tilde{M}_{\rm b\beta}(c'_1,~c'_2)^\textrm{T}=(c'_1,~c'_2)^\textrm{T}$,
    \begin{align}
	\label{Eq:Hb}
	\tilde{M}_{\rm b\beta}=&\begin{pmatrix}
	0& 	M_{\rm b\beta}^+(z_1,z_2) \\
	M_{\rm b \beta}^-(z_1,z_2)& 0
	\end{pmatrix}
    \end{align}
    where $ M_{\rm b\beta}^+(z_1,z_2)=P(z_2)/P(z_1)$, $ M_{\rm b\beta}^-(z_1,z_2) = Q(z_1)/Q(z_2)$. Due the constraint $\det[M_{\rm b \beta}]=0$ and symmetry of the boundary matrix $\mathcal{W}_+^{\rm \beta}=-\mathcal{W}_-^{\rm \beta}$, and hence the following topological winding number $\mathcal{W}_{\rm non-Bloch}^{\rm \beta}$ classifies the boundary matrix \cite{Wang_non_bloch_TI_2018, Murakami_non_bloch_TI_2019, Verma2024},
    \bea
    \label{eq:non_bloch_TI}
    \mathcal{W}_{\rm non-Bloch}^{\rm \beta}=\frac{\mathcal{W}_+^{\rm \beta}-\mathcal{W}_-^{\rm \beta}}{2}.
    \eea
    Furthermore, we find that the value of non-Bloch topological invariant $\mathcal{W}_{\rm non-Bloch}=(\mathcal{W}^{\rm P}_{\rm non-Bloch}+\mathcal{W}^{\rm H}_{\rm non-Bloch})/2$ characterize the total number of boundary states in the system. 
    
    \section{Generalized bulk-boundary correspondence}\label{sec:GBBC}
    \subsection{SPT phase.}\label{sec:BBC_SPT}
    We consider the SPT phase with $w_{\rm \alpha},\epsilon_{\alpha}=0$ with the open boundary condition. In Fig.~\ref{fig:E_W_SPT_sub_SPT} (a) and (b), we show the variation of energy spectra ($E$) and non-Bloch topological invariants ($\mathcal{W}^{\rm P}_{\rm non-Bloch}, \mathcal{W}^{\rm H}_{\rm non-Bloch},\mathcal{W}_{\rm non-Bloch}$) as a function of $u/v$, respectively. In this case, the two winding numbers $\mathcal{W}^{\rm P}_{\rm non-Bloch}, \mathcal{W}^{\rm H}_{\rm non-Bloch}$ coincide and the topological phase transitions of GBZs occur at the band touching points where the disjoint generalized momenta corresponding to two boundary modes touch GBZs for particle and hole sectors simultaneously (at $|u|=v$ in the large $N$-limit) [See Fig.~\ref{fig:OBC_TPT_SPT} and Fig.~\ref{fig:E_W_SPT_sub_SPT} (a) and (b)]. In the topologically non-trivial phase ($|u|<v$), the non-trivial values of $\mathcal{W}^{\rm P}_{\rm non-Bloch}, \mathcal{W}^{\rm H}_{\rm non-Bloch}$ are equal to the number of topological zero energy boundary modes in the system which is the same as $\mathcal{W}_{\rm non-Bloch}$ [See Fig.~\ref{fig:E_W_SPT_sub_SPT} (a) and (b)]. In Fig.~\ref{fig:E_W_SPT_sub_SPT} (a), the topological boundary states are distinguished by the inverse participation ratio (IPR), which is defined as follows,
    \begin{align}
        {\rm IPR_{n}}=\sum\limits_{i=1}^{2 N}|\psi_{n}(i)|^4/(\sum\limits_{i=1}^{2 N}|\psi_{n}(i)|^2)^2,
    \end{align}
    here $\psi_n(i)$ denote the wave function amplitude at $i$-th site for the $n$-th eigenstate $|\Psi_{n}\rangle$. The IPR distinguishes the bulk states (IPR $\sim O(1)$) and boundary states (IPR$\sim O(1/N)$).

    \subsection{Sub-SPT phase: SPT phase in the presence of global deformation.}\label{sec:BBC_sub_SPT1}
    We consider the B sub-SPT phase ($w_{\rm B}=\epsilon_{\alpha}=0$) in the open boundary condition in the presence of the global deformation term with $w_{\rm A} \neq 0$. In this case, Fig.~\ref{fig:E_W_SPT_sub_SPT} (c) and (d) show the behavior of energy spectra ($E$) and non-Bloch topological invariants ($\mathcal{W}^{\rm P}_{\rm non-Bloch}, \mathcal{W}^{\rm H}_{\rm non-Bloch},\mathcal{W}_{\rm non-Bloch}$) as a function of $u/v$. We find that two winding numbers are still quantized, and their non-trivial values correspond to the presence of boundary modes. In general, each winding number $\mathcal{W}^{\rm P}_{\rm non-Bloch}, \mathcal{W}^{\rm H}_{\rm non-Bloch}$ is different from each other. The appearance of both boundary modes is characterized by considering the winding numbers together or equivalently $\mathcal{W}_{\rm non-Bloch}$. In this case, only non-trivial values of $\mathcal{W}_{\rm non-Bloch}$ determine the total number of boundary modes. It is interesting to note here that the non-Bloch topological invariants for the particle and hole sectors coincide when the boundary modes have zero energy [Fig.~\ref{fig:E_W_SPT_sub_SPT} (c) and (d)]. Moreover, the boundary modes emerge from the bulk energy spectra at the topological phase transitions, which occur at the diabolic points (not at the band touching points) where the disjoint generalized momenta corresponding to two boundary modes touch GBZs for particle and hole sectors non-simultaneously.
    
    \begin{figure*}[htbp!]
		\centering
		\includegraphics[width=1.0\linewidth]{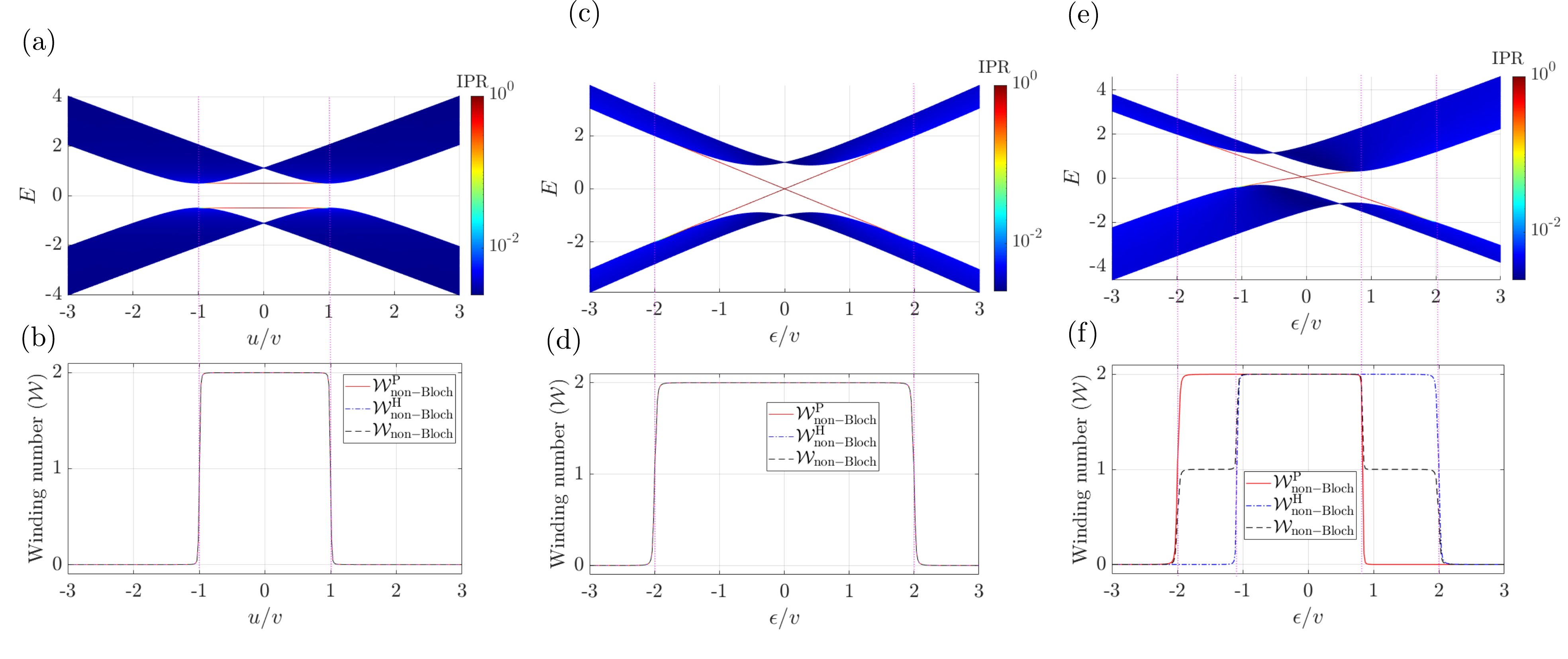}
		\caption{\textbf{Generalized bulk-boundary correspondence in the open boundary condition in the presence of symmetry-breaking mass term.} [(a) and (b)], [(c) and (d)], and [(e) and (f)] describe [the behavior of energy spectra ($E$) and non-Bloch topological invariants ($\mathcal{W}$) with system parameters] the symmetry-protected topological (SPT) phase in the presence of constant mass term, SPT phase in the presence of varying mass term, and B sub-SPT phase in the presence of varying mass term with local deformation, respectively. In this case also, the non-trivial values of $\mathcal{W}_{\rm non-Bloch}$ characterize the existence of the boundary states. IPR denotes the inverse participation ratio, which distinguishes the bulk states (IPR $\sim O(1)$) and boundary states (IPR$\sim O(1/N)$). Parameters: (a), (b): $\epsilon_{\rm A}=-\epsilon_{\rm B}=0.5,~w_{\rm \alpha}=\epsilon_{1\alpha}=\epsilon_{N\alpha}=v'=w'_{\alpha}=0$. (c), (d): $u=\epsilon,~\epsilon_{\rm A}=-\epsilon_{\rm B}=0.5\epsilon,~w_{\alpha}=\epsilon_{1\alpha}=\epsilon_{N\alpha}=v'=w'_{\alpha}=0$. (e), (f): $u=\epsilon,~\epsilon_{\rm A}=-\epsilon_{\rm B}=0.5\epsilon,~w_{\rm A}=0.4,~=\epsilon_{1\rm A}=0.1,~\epsilon_{1\rm B}=\epsilon_{N\alpha}=w_{\rm B}=\epsilon_{\alpha}=v'=w'_{\alpha}=0$. Here $\alpha\in\lbrace \rm A,B\rbrace$ and all panels have $v=1,~N=300.$}
	\label{fig:E_W_SPT_sub_SPT_mass}
    \end{figure*}
    
    \subsection{Sub-SPT phase: SPT phase in the presence of global and local deformations.}\label{sec:BBC_sub_SPT2}
    In the presence of the global ($w_{\rm A}\neq 0$) and local ($\epsilon_{1\rm A}\neq 0$) deformations, the B sub-symmetry is intact. In this case, the coefficient of the boundary equations [Eq.~\eqref{Eq_bmat} becomes $P(z_j) = -v\phi_{\rm B}^{(j)}+ (-w_{\rm A} + \epsilon_{1 \rm A} z_j) \phi_{\rm A}^{(j)},~
    Q(z_j) = -v z_j^{N+1}\phi_{\rm A}^{(j)}$. Accordingly, the paired momenta $(z_{1},~z_{2}) = (e^{i k},~e^{-i k})$ is obtained by solving following equation,
    \begin{align}\label{eqn:sol_phi_restore}
        &u v\sin[(N+1) k]+v^2\sin[N k]\nonumber\\
        &=-E(w_{\rm A}\sin[(N+1)k]-\epsilon_{1\rm A}\sin[N k]).
    \end{align}
    It is easy to observe that by varying $w_{\rm A},~\epsilon_{1\rm A}$, the right-hand side of Eq.~\eqref{eqn:sol_phi_restore} vanishes and we are left with the following equation
    \begin{align}\label{eqn:sol_phi_SPT}
        u v\sin[(N+1) k]+v^2\sin[N k]=0,
    \end{align}
    which is the same as the equation for finding paired momenta in the SPT phase. Accordingly, one can restore the doubly degenerate zero energy edge states by varying the parameters $w_{\rm A},~\epsilon_{1\rm A}$.

    In Fig.~\ref{fig:E_W_SPT_sub_SPT} (e) and (f), we show the behavior of energy spectra ($E$) and non-Bloch topological invariants ($\mathcal{W}^{\rm P}_{\rm non-Bloch}, \mathcal{W}^{\rm H}_{\rm non-Bloch},\mathcal{W}_{\rm non-Bloch}$) varying $u/v,~w_{\rm A}/v$ for a fixed value of $\epsilon_{1A}$. In this case, the two winding numbers coincide only in the presence of zero energy boundary mode, and the non-trivial values of $\mathcal{W}_{\rm non-Bloch}$ determine the total number of boundary modes in the system. Moreover, $\mathcal{W}^{\rm P}_{\rm non-Bloch}$ alone characterizes the topological phase transition of GBZ for the particle sector where the topological zero-energy mode emerges from the bulk energy spectra which is quite different and distinct from that of sub-SPT phase in the presence of the global deformation [see Sec.~\ref{sec:BBC_sub_SPT1}].
    
    \subsection{SPT and sub-SPT phases in the presence of symmetry-breaking mass term}\label{sec:BBC_mass}
    In this section, we consider the SPT and sub-SPT phases in the presence of the symmetry-breaking mass term with different on-site sublattice potentials ($\epsilon_{\rm A}=-\epsilon_{\rm B}=m$). We do not give the detailed derivation of the non-Bloch band theory here [see Appendix \ref{supp:sec:non_Bloch_theory_sub_SPT} for the full derivation]. [Fig.~\ref{fig:E_W_SPT_sub_SPT_mass} (a) and (b)], [Fig.~\ref{fig:E_W_SPT_sub_SPT_mass} (c) and (d)], and [Fig.~\ref{fig:E_W_SPT_sub_SPT_mass} (e) and (f)] describe [energy spectra ($E$) and non-Bloch topological invariants ($\mathcal{W}^{\rm P}_{\rm non-Bloch}, \mathcal{W}^{\rm H}_{\rm non-Bloch},\mathcal{W}_{\rm non-Bloch}$)] for the SPT phase in the presence of constant mass term by varying $u/v$, SPT phase in the presence of varying mass term and $u/v$, and B sub-SPT phase in the presence of varying mass term and $u/v$ with local deformation, respectively. For the SPT phase in the presence of the mass term, energies of the boundary states are only shifted by $\pm m$ whereas the topological phase transitions of GBZs coincide with that of the SPT phase [See Fig.~\ref{fig:E_W_SPT_sub_SPT} (a) and (b), Fig.~\ref{fig:E_W_SPT_sub_SPT_mass} (a) and (b), and Fig.~\ref{fig:E_W_SPT_sub_SPT_mass} (c) and (d)]. In this case, the two winding numbers coincide [Fig.~\ref{fig:E_W_SPT_sub_SPT_mass} (a) and (b)], [Fig.~\ref{fig:E_W_SPT_sub_SPT_mass} (c) and (d)]. For the sub-SPT phase in the presence of the mass term, the topological phase transitions of GBZs depend on the mass term and do not coincide with those of the sub-SPT phase [see Fig.~\ref{fig:E_W_SPT_sub_SPT} (e) and (f), Fig.~\ref{fig:E_W_SPT_sub_SPT_mass} (e) and (f)].
    In all cases, the non-trivial values of $\mathcal{W}_{\rm non-Bloch}$ characterize the total number of the boundary states.
    
    \subsection{Generalized chiral symmetric phase in the presence or absence of mass term}\label{sec:bbc_gen_chiral}
    For finite and equal next-nearest neighbor hopping $w_{\rm A}=w_{\rm B}$, the system lacks conventional chiral symmetry rather than obeys generalized chiral symmetry [see Sec. ]. In this case,
    the bulk equation is a fourth-order complex polynomial equation for a given eigenvalue $E$ with four solutions [see Appendix \ref{supp:sec:non_Bloch_theory_gen_chiral} for details of the non-Bloch band theory].
    We order these solutions as $|z_1|\leq|z_2|\leq|z_3|\leq|z_4|$. The continuum band structure for a large crystal with open boundaries \cite{Murakami_non_bloch_TI_2019} is obtained from the generalized momenta $|z_2|=|z_3|=1$, which defines the generalized Brillouin zone $\mathcal{L}_2\times \mathcal{L}_3$ of the system. Similar to the previous cases, the non-Bloch topological invariants can be defined for the ratios ($r_j,~j=1,2,3,4$) of the coefficients of outgoing ($z_2\equiv e^{-i k}$) and incoming waves ($z_3\equiv e^{i k}$), which
    will correspond to the reflection amplitude for the scattering problem [see Appendix \ref{supp:sec:non_Bloch_theory_gen_chiral} for more details]. We find that the non-Bloch topological invariants $\mathcal{W}_{\rm non-Bloch}^{(\rm P)}$ and $\mathcal{W}_{\rm non-Bloch}^{(\rm H)}$, which are defined for any of the $r_j$ with respect to the GBZs of the particle and hole sectors, successfully establish the bulk-boundary correspondence in the system.
    
    \section{Summary and discussion}\label{sec:summary} 	
In this work, we have generalized the non-Bloch band theory to describe sub-SPT phases. We have shown that the non-Bloch topological invariant can be defined as the winding number of the reflection amplitude in scattering theory. The non-Bloch topological invariant characterizes the intrinsic topology of the generalized Brillouin zone, which establishes the bulk-boundary correspondence (generalized BBC) for the SPT phases in the presence of global or local complete or partial symmetry-breaking deformations. The intrinsic topology of the GBZ applies to the generalized boundary conditions that locally break the translation symmetry of the system.

Our study can be extended to formulate the BBC in SPT or sub-SPT phases across higher dimensions, considering the influence of global or local deformation terms \cite{Poli_2017_partial_chiral, Wang2023_subsy, SSR_Schneider2023}. This investigation prompts further inquiry into whether the GBZ topology can augment the characterization and differentiation of more intricate topological phases, encompassing strong, weak, fragile, and delicate phases \cite{Bradlyn_Topo_quan_chem_2017,fragile_Vishwanath_2018,fragile_Furusaki_2021, Alexandradinata_delicate_2022}. Furthermore, the GBZ topology can corroborate the BBC for topological phases in systems devoid of internal symmetries, as illustrated by the trimer SSH model, irrespective of the presence or absence of inversion symmetry \cite{trimer_SSH_BBC_2022,trimer_SSH_BBC_2024}. 
    \\
    \section*{Acknowledgments}
    M.J.P. and S.V. acknowledge financial support from the Institute for Basic Science in the Republic of Korea through the project IBS-R024-D1. This work was supported by the research fund of Hanyang University(HY-202300000001149). This work was supported by the National Research Foundation of Korea(NRF) grant funded by the Korea government (MSIT) (Grants No. RS-2023-00252085 and No. RS-2023-00218998).





\appendix
\begin{widetext}
\section{ Classifications of generalized boundary conditions}\label{supp:sec:def_GBC}
We define the generalized boundary conditions (GBCs) as boundary conditions that break the translation symmetry locally. We consider the one-dimensional tight-binding model with two sublattices degrees of freedom, considering different on-site potentials on different sublattices and nearest and next-nearest neighbors hopping in GBCs. The corresponding model Hamiltonian ($\hat{H}=\hat{H}_{\rm bulk}+\hat{H}_{\rm d}$) can be decomposed in terms of the bulk term ($\hat{H}_{\rm bulk}$) and the boundary deformation term ($\hat{H}_{\rm d}$) as follows,
\begin{align}\label{supp:general_model_ham}
	\hat{H}_{\rm bulk}&=\hat{H}_{\rm SSH}+\sum_{\alpha=\rm A,B}\hat{H}_{\rm \alpha\alpha},\\
 \hat{H}_{\rm SSH}&=\sum_{j=1}^{N}
    \big[ u \hat{c}^\dagger_{j \rm B}\hat{c}_{j \rm A}
    +v(1-\delta_{j N}) \hat{c}^\dagger_{(j+1) \rm A}\hat{c}_{j \rm B} + {\rm h.c.}\big],~
    \hat{H}_{\alpha\alpha}=\sum_{j=1}^{N}\big[
     \epsilon_{\alpha} \hat{c}^\dagger_{j\alpha}\hat{c}_{j\alpha}
    + w_{\alpha} (1-\delta_{j N}) \hat{c}^\dagger_{(j+1)\alpha}\hat{c}_{j\alpha}
    +{\rm h.c.}\big],\nonumber\\
    \hat{H}_{\rm d}&=
    v'\hat{c}^\dagger_{1 \rm A}\hat{c}_{N \rm B}+v' \hat{c}^\dagger_{N \rm B}\hat{c}_{1 \rm A}+\sum_{\alpha={\rm A,B}}[\epsilon_{1\alpha} \hat{c}^{\dagger}_{1\alpha} \hat{c}_{1\alpha} +\epsilon_{N\alpha} \hat{c}^\dagger_{N\alpha}\hat{c}_{N\alpha} + w'_{\alpha}\hat{c}^\dagger_{1\alpha}\hat{c}_{N\alpha}+w'_{\alpha} \hat{c}^\dagger_{N\alpha}\hat{c}_{1\alpha}].
    \end{align}
	where $|j\rangle \equiv \hat{c}^{\dagger}_j|0\rangle$ is the localized state at $j$-th site. $N$-denotes the total number of unit cells consisting of A and B sublattices [See Fig.~\ref{fig:tight_binding_GBC}]. $u$ ($v$) denotes the nearest neighbor intra-cell (inter-cell) hopping. $w_{\rm A}$ ($w_{\rm B}$) denotes the next-nearest neighbor hopping for A (B) sublattices. $\epsilon_{\rm A}$ ($\epsilon_{\rm B}$) denotes the on-site potentials for A (B) sublattices. In a general setting, different combinations of boundary parameters ($v', w'_{\rm \alpha}, \epsilon_{1\alpha},\epsilon_{N\alpha}$) with $\alpha\in\lbrace\rm A, B\rbrace$ describe all possible sets of GBC [Fig.~\ref{fig:tight_binding_GBC}]. For example, $(v', w_{\rm \alpha},\epsilon_{1\alpha},\epsilon_{N\alpha}) =(v,w_{\alpha},0,0)$ corresponds to the periodic boundary condition (PBC), while $(v', w_{\rm \alpha},\epsilon_{1\alpha},\epsilon_{N\alpha}) =(0,0,0,0,0)$ corresponds to the open boundary condition (OBC). Table~\ref{tab:def_GBC} summarizes the definitions of the different types of boundary conditions.
 
    \begin{center}
	\begin{table}[h]
    \def\arraystretch{1.5}
    \begin{tabular}{|l|l|}
     \hline
    Periodic boundary condition (PBC) & $v'=v,~w'_{\rm\alpha}=w_{\rm\alpha},$ and $\epsilon_{1\alpha}=\epsilon_{N\alpha}=0$ \\
    \hline
    Generalized boundary conditions (GBC) & All possible combinations of $t'_{2},~t^{\prime\rm\alpha}_{2},~\epsilon_{1\alpha},~\epsilon_{N\alpha}$\\
    (i) Open boundary condition (OBC) & $v'=0,~w'_{\rm\alpha}=0,$ and $\epsilon_{1\alpha}=\epsilon_{N\alpha}=0$ \\
    (ii) Generalized boundary conditions-(I) (GBC-(I)) &$v'=0,~w'_{\rm\alpha}=0$, and $\epsilon_{1\alpha},~\epsilon_{N\alpha}\neq 0$ \\
    (iii) Generalized boundary conditions-(II) (GBC-(II)) & $v'=v,~w'_{\rm\alpha}=w_{\rm\alpha}$, and $\epsilon_{1\alpha},~\epsilon_{N\alpha}\neq0$ \\
    (iv) Generalized boundary conditions-(III) (GBC-(III)) &  $v'<v$, $w'_{\rm\alpha}<w_{\rm\alpha}$, and 
    $\epsilon_{1\alpha},~\epsilon_{N\alpha}=0$ \\
    (v) Generalized boundary conditions-(IV) (GBC-(IV)) & $v'>v$, 
    $w'_{\rm\alpha}>w_{\rm\alpha}$, and $\epsilon_{1\alpha},~\epsilon_{N\alpha}=0$ \\
    \hline
    \end{tabular}
    \caption{Definitions of different types of generalized boundary conditions.}\label{tab:def_GBC}
    \end{table}
    \end{center}

    \section{Detailed methods of non-Bloch band theory }\label{supp:sec:non_Bloch_theory_g}
    In this section, we describe the detailed methods of the non-Bloch band theory \cite{Murakami_non_bloch_TI_2019, chenGBC2021} to solve the eigenvalue problem of $\hat{H}|\Psi\rangle=E|\Psi\rangle$ in the generalized boundary conditions. The eigenstate, $|\Psi\rangle = \sum_{n=1}^{N}(\psi_{n \rm A}\hat{c}^{\dagger}_{n \rm A}+\psi_{n \rm B}\hat{c}^{\dagger}_{n \rm B})|0\rangle$, with the eigenvalue $E$ is required to satisfy the bulk and the boundary equations simultaneously.\\ 
    \textbf{Bulk equation}: The bulk equations are given as,
    \begin{align}\label{supp:eq:bulk_eqn_g}
	&w_{\rm A}\psi_{(n-1) \rm A}+v\psi_{(n-1) \rm B}+(\epsilon_{\rm A}-E)\psi_{n \rm A}+u\psi_{n         \rm B}+w_{\rm A}\psi_{(n+1) \rm A}=0\nonumber\\
        &w_{\rm B}\psi_{(n-1) \rm B}+u\psi_{n \rm A}+(\epsilon_{\rm B}-E)\psi_{n \rm B}+v\psi_{(n+1)\rm A}+w_{\rm B}\psi_{(n+1)\rm B}=0~~{\rm for}~~n\in\lbrace 2,...,N-1\rbrace.
    \end{align}
    We solve the bulk equations using single wave vector ansatz of the wave function, $\psi_{n\alpha} \propto \phi_{\alpha}^{(j)}z_j^n,~\alpha\in\lbrace\rm A, B\rbrace$ which respects the translation symmetry of the bulk. The bulk equations transform as follows,
    \bea 
    \label{eq:bulk_eqn_2_g}
    \frac{\phi_{\rm A}^{(j)}}{\phi_{\rm B}^{(j)}}=-\frac{\frac{v}{z_j}+u}{\frac{w_{\rm A}}{z_j} +(\epsilon_{\rm A}-E) + w_{\rm A}z_j}=-\frac{\frac{w_{\rm B}}{z_j} +(\epsilon_{\rm B}-E) + w_{\rm B}z_j}{v z_j+u},
    \eea
    which immediately leads to,
	\bea 
        \label{supp:bulk_poly_g}
	(\epsilon_{\rm A}-E)(\epsilon_{\rm B}-E) &+& (\epsilon_{\rm A}-E)\Big(\frac{w_{\rm B}}{z_j}+w_{\rm B}z_j\Big) + (\epsilon_{\rm B}-E)\Big(\frac{w_{\rm A}}{z_j}+w_{\rm A}z_j\Big)\nonumber\\ 
 &+& \Big(\frac{w_{\rm A}}{z_j}+w_{\rm A}z_j\Big)\Big(\frac{w_{\rm B}}{z_j}+w_{\rm B}z_j\Big)-\Big(\frac{v}{z_j}+u\Big)(v z_j+u)= 0.
	\eea 
    The above equation is a fourth-order complex polynomial equation for a given eigenvalue $E$. In general, the linear superposition of the four ansatz forms general solutions of the bulk equations, 
    \bea
	\label{supp:bulk_ansatz_g}
	\psi_{n\alpha}=c_1  \phi_{\alpha}^{(1)} z_1^n +c_2 \phi_{\alpha}^{(2)} z_2^n  +c_3  \phi_{\alpha}^{(3)} z_3^n + c_4  \phi_{\alpha}^{(4)} z_4^n,~\alpha\in\lbrace\rm A,B\rbrace.
    \eea
    \textbf{Boundary equation}: The boundary equations are given as,
    \begin{align}\label{eq:boundary_eqn_gg_i}
	w_{\rm A} \psi_{N \rm A}+ v^{\prime} \psi_{N \rm B} + (\epsilon_{1 \rm A}+\epsilon_{\rm A} - E) \psi_{1 \rm A} + u \psi_{1 \rm B} +w_{\rm A} \psi_{2 \rm A} &= 0, \\
    w_{\rm B} \psi_{2 \rm B}+ v \psi_{2 \rm A} + (\epsilon_{1 \rm B}+\epsilon_{\rm B} - E) \psi_{1 \rm B} + u \psi_{1 \rm A} +w'_{ \rm B} \psi_{N \rm B} &= 0, \\
    w'_{\rm A}\psi_{1 \rm A} + u \psi_{N \rm B}+ (\epsilon_{N \rm A}+ \epsilon_{\rm A} - E) \psi_{N \rm A} + v \psi_{(N-1) \rm B} + w_{\rm A} \psi_{(N-1) \rm A}& = 0 ,\\
	w'_{\rm B}\psi_{1 \rm B} + v^{\prime} \psi_{1 \rm A} + (\epsilon_{N \rm B} +\epsilon_{\rm B}- E) \psi_{N \rm B} + u \psi_{N \rm A} + w_{\rm B} \psi_{(N-1) \rm B}& = 0,
    \end{align}
    which are further simplified to the following form with the help of bulk equations,
    \begin{align}\label{eq:boundary_eqn_gg_f}
	w'_{ \rm A} \psi_{N \rm A}+ v^{\prime} \psi_{N \rm B} + \epsilon_{1 \rm A} \psi_{1 \rm A} - v \psi_{0 \rm B} -w_{\rm A} \psi_{0 \rm A} &= 0, \\
    -w_{\rm B} \psi_{0 \rm B}+ \epsilon_{1 \rm B}  \psi_{1 \rm B}  +w'_{ \rm B} \psi_{N \rm B} &= 0, \\
    w'_{ \rm A}\psi_{1 \rm A} + \epsilon_{N \rm A} \psi_{N \rm A} - w_{\rm A} \psi_{(N+1) \rm A}& = 0 ,\\
	w'_{ \rm B}\psi_{1 \rm B} + v^{\prime} \psi_{1 \rm A} + \epsilon_{N \rm B}  \psi_{N \rm B} - v \psi_{(N+1) \rm A} - w_{\rm B} \psi_{(N+1 \rm B}& = 0 .
    \end{align}
    By plugging in the bulk ansatz in Eq. \eqref{supp:bulk_ansatz_g}, the boundary equation [Eq. \eqref{eq:boundary_eqn_gg_f}] can be compactly rewritten as the form of the eigenvalue equation,
	\begin{align}
	\label{eqs:H_b_g}
	M_{\rm b} \begin{pmatrix}
	c_1\\
	c_2\\
        c_3\\
        c_4\end{pmatrix} =0, ~~{\rm with}~~ M_{\rm b} = \begin{pmatrix}
	P_1(z_1)& P_1(z_2) &P_1(z_3) & P_1(z_4) \\
        Q_1(z_1)& Q_1(z_2) &Q_1(z_3) & Q_1(z_4) \\
        P_N(z_1)& P_N (z_2) & P_N (z_3) & P_N(z_4) \\
	Q_N(z_1)& Q_N(z_2) &Q_N(z_3) & Q_N(z_4) 
	\end{pmatrix},
	\end{align}
	where 
    \begin{align}
    P_1(z_j) &= (-v+v^{\prime} z_j^{N})\phi_{\rm B}^{(j)}+ (-w_{\rm A} + \epsilon_{1 \rm A} z_j+ w'_{\rm A}z_j^N) \phi_{\rm A}^{(j)}, \\
    Q_1(z_j) &= (-w_{\rm B}+\epsilon_{1 \rm B}z_j +w'_{\rm B} z_j^{N})\phi_{\rm B}^{(j)}, \\
    P_N(z_j) &=  (-w_{\rm A}z_j^{N+1}+\epsilon_{N \rm A}z_j^N +w'_{ \rm A} z_j)\phi_{\rm A}^{(j)},\\
    Q_N(z_j) &= (-v z_j^{N+1}+v^{\prime} z_j)\phi_{\rm A}^{(j)}+ (-w_{\rm B}z_j^{N+1}+\epsilon_{N \rm B} z_j^{N} + w'_{\rm B}z_j)\phi_{\rm B}^{(j)},
    \end{align}
    Here $\phi_{\rm A}^{(j)}$ and $\phi_{\rm B}^{(j)}$ are related via Eq.~\eqref{eq:bulk_eqn_2_g}.

    \textbf{Eigenstate Solutions and generalized Brillouin zones:}	 The non-zero solutions are obtained by solving the equation, $\det[M_{\rm b}] =0$ together with the constraint that all $(z_{1n},z_{2n},z_{3n}, z_{4n})$'s with $n
    \in\lbrace 1,...,N\rbrace$ should satisfy the bulk equations [Eq.~\eqref{supp:bulk_poly_g}] as well. Consequently, we obtain all the eigenvalues and corresponding eigenstates from Eq.~\ref{supp:bulk_poly_g} and Eq.~\ref{supp:bulk_ansatz_g}. The contours of generalized momenta define the generalized Brillouin zones (GBZs), whereas the disjoint generalized momenta inside (outside) the GBZs describe the evanescent waves localized at the system's boundary. In principle, one can get the non-zero solution of the boundary equation [Eq. \eqref{eqs:H_b_g}] via numerical means. However, in this case, it is quite cumbersome to proceed analytically. Therefore, for simplicity, we first consider some special cases, for example, the symmetry-protected, sub-symmetry-protected phases, which are described by the Hamiltonians $\hat{H}_{\rm SSH}$, $\hat{H}_{\rm SSH}+\hat{H}_{\rm AA}$ or $\hat{H}_{\rm SSH}+\hat{H}_{\rm BB}$, respectively (see the main text for the definition of symmetry-protected and sub-symmetry-protected phases).

    \section{Sub-symmetry-protected topological phase in the presence of the mass term}\label{supp:sec:non_Bloch_theory_sub_SPT}
	
	In this section, we elaborate on the non-Bloch band theory \cite{Murakami_non_bloch_TI_2019, chenGBC2021} for the sub-symmetry-protected phase with model Hamiltonian $\hat{H}_{\rm SSH}+\hat{H}_{ \rm AA}$. The calculations are similar and straightforward for the other sub-symmetry-protected phase described by $\hat{H}_{\rm SSH}+\hat{H}_{\rm BB}$. Moreover, one can obtain the generalized Brillouin zones, band structures, and topological invariants for the symmetry-protected phase described by $H_{\rm SSH}$ by setting $w_{\rm A} =0$ throughout this section. The eigenstate $|\Psi\rangle = \sum_{n=1}^{N}(\psi_{n \rm A}\hat{c}^{\dagger}_{n \rm A}+\psi_{n \rm B}\hat{c}^{\dagger}_{n \rm B})|0\rangle$, with the eigenvalue $E$ is obtained via solving the bulk and the boundary equations simultaneously.
 
    \textbf{Bulk equations}: The bulk equations [Eq.~\ref{supp:eq:bulk_eqn_g}] reduces to,
	\begin{align}\label{supp:eq:bulk_eqn_sub}
	&w_{\rm A}\psi_{(n-1) \rm A}+v\psi_{(n-1) \rm B}+(\epsilon_{\rm A}-E)\psi_{n \rm A}+u\psi_{n \rm B}+w_{\rm A}\psi_{(n+1) \rm A}=0,\nonumber\\
        &u\psi_{n \rm A}+(\epsilon_{\rm B}-E)\psi_{n \rm B}+v\psi_{(n+1) \rm A}=0~~{\rm for}~~n\in\lbrace 2,...,N-1\rbrace.
	\end{align}
    Again using the single wave vector ansatz of the wave function, $\psi_{n\alpha} \propto \phi_{\alpha}^{(j)}z_j^n,~\alpha\in\lbrace\rm A, B\rbrace$, the bulk equation transforms as,
    \bea 
    \label{supp:eq:bulk_eqn_sub_2}
    \frac{\phi_{\rm A}^{(j)}}{\phi_{\rm B}^{(j)}}=-\frac{\frac{v}{z_j}+u}{\frac{w_{\rm A}}{z_j} +(\epsilon_{\rm A} - E) + w_{\rm A}z_j}=-\frac{(\epsilon_{\rm B} - E) }{ v z_j + u},
    \eea
    which immediately leads to,
	\bea 
        \label{supp:eq:bulk_eqn_poly_sub}
	(\epsilon_{\rm A} - E) (\epsilon_{\rm B} -E)+(\epsilon_{\rm B} - E) \Big(\frac{w_{\rm A}}{z_j}+w_{\rm A}z_j\Big) -\Big(\frac{v}{z_j}+u\Big)(v z_j+u)= 0.
	\eea 
    This second-order complex polynomial equation has two solutions $(z_1,~z_2)$ for a given eigenvalue $E$. In this case, the eigenvalues take the following form,
    \bea 
        \label{supp:eq:eigenvalue_sub}
	E_{\pm}=\frac{1}{2}\Big(\epsilon_{\rm A} +\epsilon_{\rm B} + \frac{w_{\rm A}}{z_j}+w_{\rm A} z_j\Big)\pm\frac{1}{2}\sqrt{\Big(\epsilon_{\rm A} - \epsilon_{\rm B} + \frac{w_{\rm A}}{z_j}+w_{\rm A}z_j\Big)^2+4\Big(\frac{v}{z_j}+u\Big)(v z_j+u)\Big)}.
	\eea 
    In general, the linear superposition of the two ansatz forms general solutions of the bulk equations, 
	\bea
	\label{supp:eq:bulk_ansatz_sub}
	\psi_{n\alpha}=c_1  \phi_{\alpha}^{(1)} z_1^n +c_2 \phi_{\alpha}^{(2)} z_2^n ,~\alpha\in\lbrace\rm A,B\rbrace.
	\eea
	
    \textbf{Boundary equations}: In this case, the boundary equations [Eq.~\eqref{eq:boundary_eqn_gg_i}] are significantly simplified and becomes 
    \begin{align}\label{eq:boundary_eqn_sub_i}
	w'_{\rm A} \psi_{N \rm A}+ v^{\prime} \psi_{N \rm B} + (\epsilon_{1 \rm  A} + \epsilon_{\rm A} - E_{\pm}) \psi_{1 \rm A} + u \psi_{1 \rm B} +w_{\rm A} \psi_{2 \rm A} &= 0, \\
	 v^{\prime} \psi_{1 \rm A}  + (\epsilon_{N \rm  B} + \epsilon_{\rm B} - E_{\pm}) \psi_{N \rm B} + u \psi_{N \rm A} & = 0 .
    \end{align}
    which are further simplified with the help of bulk equations and  take the following form,
    \begin{align}\label{supp:eq:boundary_eqn_sub_f}
	w'_{\rm A} \psi_{N \rm A}+ v^{\prime} \psi_{N \rm B} + \epsilon_{1 \rm A} \psi_{1 \rm A} - v   \psi_{0 \rm B} -w_{\rm A} \psi_{0 \rm A} &= 0, \nonumber\\
	v^{\prime} \psi_{1 \rm A} + \epsilon_{N \rm  B} \psi_{N \rm B} - v \psi_{(N+1) \rm A} & = 0 .
    \end{align}
    By plugging in the bulk ansatz in Eq. \eqref{supp:eq:bulk_ansatz_sub}, the boundary equation [Eq. \eqref{supp:eq:boundary_eqn_sub_f}] can be compactly rewritten as the form of the eigenvalue equation,
    \begin{align}
	\label{eqs:H_b_sub}
	M_{\rm b} \begin{pmatrix}
	c_1\\
	c_2
 \end{pmatrix} =0, ~~{\rm with}~~ M_{\rm b} = \begin{pmatrix}
	P(z_1)& P(z_2)  \\
    Q(z_1)& Q(z_2)
	\end{pmatrix},
    \end{align}
	where 
    \begin{align}
    P(z_j) &= (-v+v^{\prime} z_j^{N})\phi_{\rm B}^{(j)}+ (-w_{\rm A} + \epsilon_{1 \rm A} z_j+ w'_{\rm A}z_j^N) \phi_{\rm A}^{(j)}, \\
    Q(z_j) &= \epsilon_{N \rm B} z_j^{N}\phi_{\rm B}^{(j)} + (-v z_j^{N+1}+v^{\prime} z_i)\phi_{\rm A}^{(j)},
    \end{align}
    Here $\phi_{\rm A}^{(j)}$ and $\phi_{\rm B}^{(j)}$ are related via bulk equations [Eq.~\eqref{supp:eq:bulk_eqn_sub_2}].
    
    \textbf{Eigenstate Solutions and generalized Brillouin zones:}	The non-zero solutions of the boundary equation [Eq. \eqref{eqs:H_b_sub}] are obtained by solving for the equation, $\det[M_{\rm b}] =0$, which takes the explicit form as follows,
    \begin{align}\label{supp:eq:sol_sub_z}
	&(z_2^{N+1}-z_1^{N+1})+\Bigg(\frac{v}{u}-\frac{(v')^2}{uv} + \frac{\epsilon_{1 \rm A}\epsilon_{N\rm B}}{uv} -  \frac{\epsilon_{N\rm B}w_{\rm A}}{ v^2} \Bigg)(z_2^N-z_1^N) + \Bigg(
	-\Big(\frac{v'}
	{v}\Big)^2 + \frac{\epsilon_{1 \rm A}\epsilon_{N\rm B}}{v^2} +  \frac{\epsilon_{N\rm B}w_{\rm A}}{ uv}\Bigg)(z_2^{N-1}-z_1^{N-1})\nonumber\\
 &+\Bigg(-2\frac{v^{\prime}}{v}+\frac{\epsilon_{N\rm B}w'_{\rm A}}{ u v}\Bigg)(z_2-z_1)=-\frac{\epsilon_{N\rm B}(\epsilon_{\rm A}-E_{\pm})}{uv}(z_2^N-z_1^N)\nonumber\\&+\Bigg(\frac{(\epsilon_{\rm B}-E_{\pm})}
	{uv}\Bigg)\Big[w_{\rm A}(z_2^{N+1}-z_1^{N+1})-\epsilon_{1 \rm A}(z_2^N-z_1^N) + \Bigg(
	\frac{v'w'_{\rm A}}
	{v}\Bigg)(z_2^{N-1}-z_1^{N-1})-
	\Bigg(\frac{v^{\prime}w_{\rm A}}{v}+\frac{t_{2}w'_{\rm A}}{v}\Bigg)(z_2-z_1)\Big].
    \end{align}
    Moreover, irrespective of the eigenvalue $E$, the two generalized momenta $(z_1,~z_2)$ are paired with complementary condition $z_1 z_2 =r=1$. The paired momenta satisfying $z_1 z_2=1$ can be expressed in the polar coordinates as follows,
    \begin{align}
        \label{eqn_for_r}
        z_1 = e^{ik}, ~~z_2 = e^{-ik}, 
    \end{align}
    where $k_n\equiv k\in\mathbb{C}$ with $n\in\lbrace 1,..., N\rbrace$ can be a complex number. The contours of two paired momenta define the GBZs ($\mathcal{L}_1\times\mathcal{L}_2$) which in the present case coincide $|z_1|=|z_2|=1$. The disjoint momenta inside (outside) GBZs describes the boundary modes of the system. The eigenvalues are further simplified to the following form,
    \bea
    \label{supp:eq:eigenvalue_sub_k}
    E_{\pm}(k)\equiv E_{\pm}=\frac{\epsilon_{\rm A} +\epsilon_{\rm B}}{2}+ w_{\rm A}\cos[k]\pm\sqrt{\Big(\frac{\epsilon_{\rm A} -\epsilon_{\rm B}}{2}+w_{\rm A}\cos[k]\Big)^2 + u^2+v^2+2 u v \cos[k]}.
    \eea 
    Accordingly, Eq.~\eqref{supp:eq:sol_sub_z} can be rewritten as the following form,
    \begin{align}\label{supp:eq:sol_sub_k}
    &\eta_0\sin[(N+1)k]+\eta_1\sin[Nk]+\eta_2\sin[(N-1)k]+\eta_3\sin[k]\nonumber\\&=(\epsilon_{\rm B}-E_{\pm})
	(\eta'_0\sin[(N+1)k]+\eta'_1\sin[Nk] + \eta'_2\sin[(N-1)k]+
	\eta'_3\sin[k]) +(\epsilon_{\rm A}-E_{\pm})
	\eta''_1\sin[Nk],
    \end{align}
	where
    \begin{align}
        \eta_0&=1,~~
	\eta_1 = \Bigg(\frac{v}{u}-\frac{(v')^2}{u v} + \frac{\epsilon_{1 \rm A}\epsilon_{N\rm B}}{u v} -  \frac{\epsilon_{N\rm B}w_{\rm A}}{ v^2} \Bigg),~~\eta_2 = \Bigg(
	-\Big(\frac{v'}
	{v}\Big)^2 + \frac{\epsilon_{1 \rm A}\epsilon_{N\rm B}}{v^2} +  \frac{\epsilon_{N\rm B}w_{\rm A}}{ u v}\Bigg),~~\eta_3 = \Bigg(-2\frac{v^{\prime}}{v}+\frac{\epsilon_{N\rm B}w'_{\rm A}}{ u v}\Bigg),\nonumber\\
 \eta'_0&=\frac{w_{\rm A}}{u v},~~\eta'_{1} = -\Bigg(\frac{\epsilon_{1 \rm A}}{u v}\Bigg),
	\eta''_{1} = -\Bigg(\frac{\epsilon_{N \rm B}}{u v}\Bigg),~~\eta'_2 = \Bigg(
	\frac{v'w'_{\rm A}}
	{u v^2}\Bigg),~~\eta'_3 = -
	\Bigg(\frac{v^{\prime}w_{\rm A}}{u v^2}+\frac{v w'_{\rm A}}{u v^2}\Bigg),
    \end{align}
    In practice, we solve Eq.~\eqref{supp:eq:sol_sub_k} to determine the spectral properties of the system.

    \begin{figure*}[htbp!]
		\centering
		\includegraphics[width=1\linewidth]{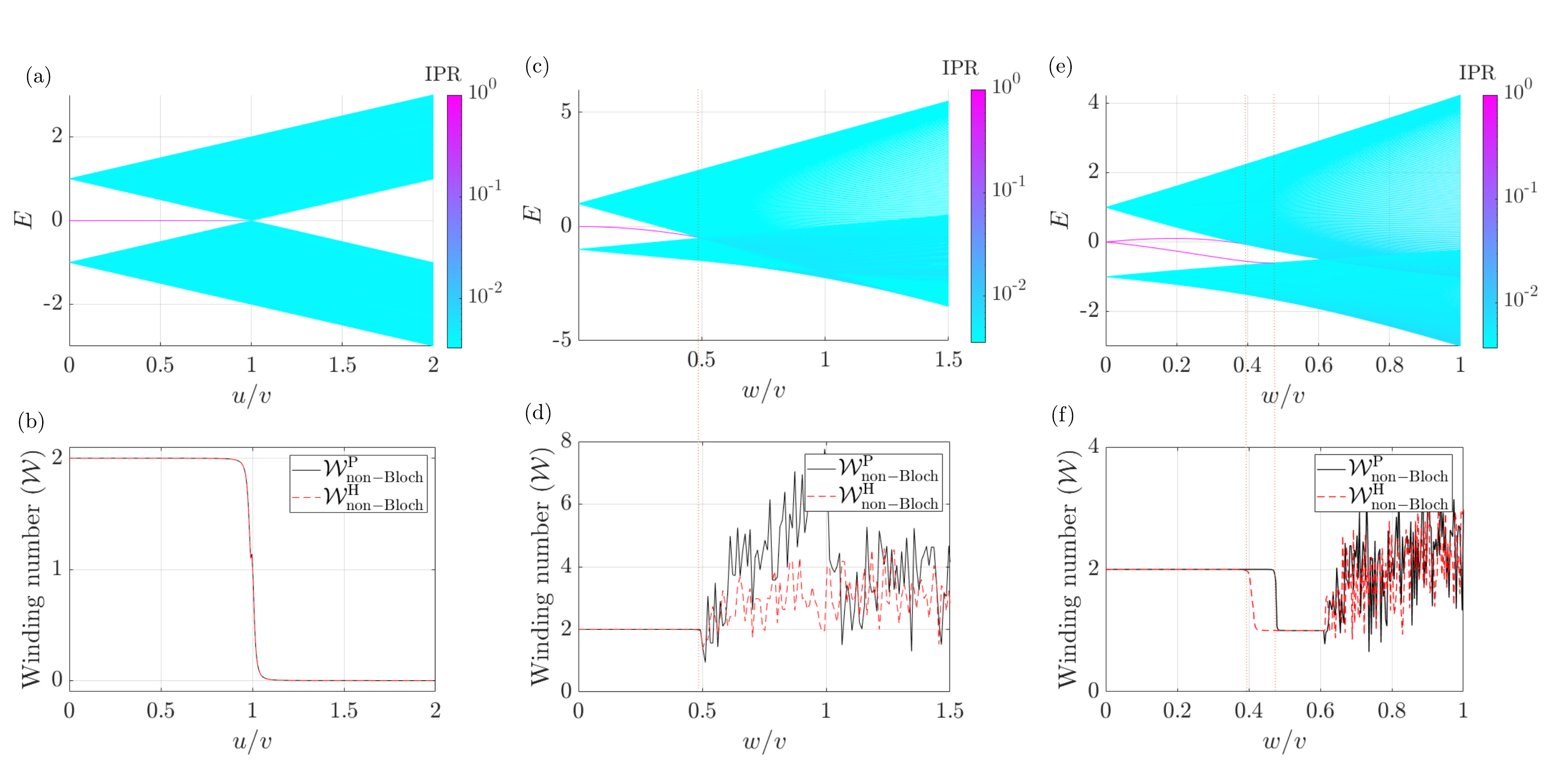}
		\caption{\textbf{Generalized bulk-boundary correspondence in the open boundary condition for generalized chiral symmetric system.} [(a) and (b)], [(c) and (d)], and [(e) and (f)] show [the variation of energy spectra ($E$) and non-Bloch topological invariants ($\mathcal{W}$) with system parameters] in the symmetry-protected topological (SPT) phase, generalized chiral symmetric phase, generalized chiral symmetric phase with mass term, respectively. The quantized changes in $\mathcal{W}_{\rm non-Bloch}$ characterize the existence of the boundary states. IPR denotes the inverse participation ratio, which distinguishes the bulk states (IPR $\sim O(1)$) and boundary states (IPR$\sim O(1/N)$). Parameters: (a), (b): $w_{\alpha}=\epsilon_{1\rm \alpha}=\epsilon_{N\alpha}=\epsilon_{\alpha}=v'=w'_{\alpha}=0$. (c), (d): $u=w_{\rm A}=w_{\rm B}\equiv w$,~$\epsilon_{1\rm \alpha}=\epsilon_{N\alpha}=\epsilon_{\alpha}=v'=w'_{\alpha}=0$. (e), (f): $u=w_{\rm A}=w_{\rm B}=\epsilon_{\rm A}=-\epsilon_{\rm B}\equiv w$,~$\epsilon_{1\rm \alpha}=\epsilon_{N\alpha}=v'=w'_{\alpha}=0$. Here $\alpha\in\lbrace \rm A, B\rbrace$ and all panels have $v=1,~N=400.$}
	\label{fig:E_W_generalized_chiral}
    \end{figure*}
    
    \section{Generalized chiral symmetric phases in the presence of the mass term}\label{supp:sec:non_Bloch_theory_gen_chiral}
    In this section, we discuss the non-Bloch band theory \cite{Murakami_non_bloch_TI_2019, chenGBC2021} for generalized chiral symmetric phases in the presence or absence of the mass term in the open boundary condition. The bulk equations are the same as Eq.~\eqref{supp:eq:bulk_eqn_g}, a fourth-order complex polynomial equation for a given eigenvalue $E$ with four solutions. These solutions satisfy the following constraints due to Eq.~\eqref{supp:bulk_poly_g} 
    \begin{align}
        z_1+z_2+z_3+z_3 =z_1 z_2 z_3+z_2 z_3 z_4 + z_3 z_4 z_1+z_4 z_1 z_2&= -((\epsilon_{\rm A}-E)w_{\rm B}+(\epsilon_{\rm B}-E)w_{\rm A}-u v)/(w_{\rm A}w_{\rm B})\\
        z_1 z_2 +z_2 z_3 +z_3 z_4 +z_4 z_1 + z_4 z_3 +z_3 z_1&= ((\epsilon_{\rm A}-E)(\epsilon_{\rm B}-E) + 2 w_{\rm A}w_{\rm B}-u^2-v^2)/(w_{\rm A}w_{\rm B})\\
        z_1 z_2 z_3 z_4 &= 1.
    \end{align}
    Accordingly, the solutions must satisfy $z_1 z_4=1$ and $z_2 z_3 =1$. Writing $(z_2,z_3)=(e^{-i k}, e^{i k})$ further simplifies the eigenvalue equations to the following form,
    \begin{align}
        E_{\pm} =\frac{1}{2}(\gamma^2\pm\sqrt{b^2-4\gamma\delta},~ \gamma=(\epsilon_{\rm A}+\epsilon_{\rm B}+2(w_{\rm A}+w_{\rm B})\cos(k)),\nonumber\\
        \delta=2(\epsilon_{\rm A}w_{\rm B}+\epsilon_{\rm B}w_{\rm A})\cos(k) +\epsilon_{\rm A}\epsilon_{\rm B}+ 4 w_{\rm A} w_{\rm B} \cos^2(k)-v^2-u^2-2uv\cos(k).
    \end{align}

    The boundary equations [Eq.~\eqref{eqs:H_b_g}] are modified with following coefficients of the boundary matrix $M_{b}$
    \begin{align}
    P_1(z_j) &= -v\phi_{\rm B}^{(j)}-w_{\rm A} \phi_{\rm A}^{(j)}, \\
    Q_1(z_j) &= -w_{\rm B}\phi_{\rm B}^{(j)}, \\
    P_N(z_j) &=  -w_{\rm A}z_j^{N+1}\phi_{\rm A}^{(j)},\\
    Q_N(z_j) &= -v z_j^{N+1}\phi_{\rm A}^{(j)}-w_{\rm B}z_j^{N+1}\phi_{\rm B}^{(j)},
    \end{align}
    We order the non-zero solutions of $\det[M_{\rm b}]=0$ as $|z_1|\leq|z_2|\leq|z_3|\leq|z_4|$. The continuum band structure for a large crystal with open boundaries \cite{Murakami_non_bloch_TI_2019} is obtained from the generalized momenta $|z_2|=|z_3|=1$, which defines the generalized Brillouin zone $\mathcal{L}_2\times \mathcal{L}_3$ of the system. The GBZ is further divided into the particle ($\mathcal{L}_2^{(\rm P)}\times \mathcal{L}_3^{(\rm P)}$ for $E_{+}$) and hole ($\mathcal{L}_2^{(\rm H)}\times \mathcal{L}_3^{(\rm H)}$ for $E_{-}$) sectors. However, the bound states are determined by the disjoint generalized momenta.

    \textbf{Topological invariants.} In this case, it is not entirely clear how to define the topological invariant since there are four generalized momenta and associated coefficients. However, similar to the previous case, one can think of $c'_2\equiv c_2\phi_{\alpha}^{(2)},~c'_3\equiv \phi_{\alpha}^{(3)}$ as the coefficients of outgoing ($z_2\equiv e^{-i k}$) and incoming waves ($z_3\equiv e^{i k}$) in the full wave function $\psi_{n\alpha}=c_1  \phi_{\alpha}^{(1)} z_1^n +c_2 \phi_{\alpha}^{(2)} z_2^n  +c_3  \phi_{\alpha}^{(3)} z_3^n + c_4  \phi_{\alpha}^{(4)} z_4^n$~\cite{Verma2024}. Accordingly, the meromorphic functions 
    \begin{align}
    r_1&\equiv-\frac{c'_2}{c'_3}=\frac{P_1(z_3)\phi_{\alpha}^{(2)}}{P_1(z_2)\phi_{\alpha}^{(3)}}=\frac{v\phi_{\rm B}^{(3)}/\phi_{\rm A}^{(3)}+w_{\rm A}}{v\phi_{\rm B}^{(2)}/\phi_{\rm A}^{(2)}+w_{\rm A}}\\
    r_2&\equiv-\frac{c'_2}{c'_3}=\frac{Q_1(z_3)\phi_{\alpha}^{(2)}}{Q_1(z_2)\phi_{\alpha}^{(3)}}=\frac{\phi_{\rm B}^{(3)}/\phi_{\rm A}^{(3)}}{\phi_{\rm B}^{(2)}/\phi_{\rm A}^{(2)}}=\frac{v z_3 +u}{v z_2 +u}\\
    r_3&\equiv-\frac{c'_2}{c'_3}=\frac{P_N(z_3)\phi_{\alpha}^{(2)}}{P_N(z_2)\phi_{\alpha}^{(3)}}=\frac{z_3^{N+1}}{z_2^{N+1}}\\
    r_4&\equiv-\frac{c'_2}{c'_3}=\frac{Q_N(z_3)\phi_{\alpha}^{(2)}}{Q_N(z_2)\phi_{\alpha}^{(3)}}=\frac{z_3^{N+1}}{z_2^{N+1}}\frac{v+\phi_{\rm B}^{(3)}/\phi_{\rm A}^{(3)}w_{\rm B}}{v+\phi_{\rm B}^{(2)}/\phi_{\rm A}^{(2)}w_{\rm B}}
    \end{align} will correspond to the reflection amplitude for the scattering problem. We find that the non-Bloch topological invariants $\mathcal{W}_{\rm non-Bloch}^{(\rm P)}$ and $\mathcal{W}_{\rm non-Bloch}^{(\rm H)}$, which are defined for any of the above meromorphic functions with respect to the GBZs of the particle and hole sectors, successfully establish the bulk-boundary correspondence in the system [see Fig.~\ref{fig:E_W_generalized_chiral}]. However, further investigation is needed to understand the symmetries of the boundary matrix, which guarantees the topological invariants' quantization, and the relation of the total number of boundary modes to the topological invariants. 
    
    \end{widetext}

\bibliography{reference}
    
\end{document}